\documentclass[epj,nopacs]{svjour}
\usepackage{graphics}
\usepackage{graphicx}
\usepackage{epsfig}
\usepackage{units}
\usepackage{cite}
\usepackage[switch]{lineno}
\usepackage{calc}
\usepackage{endnotes}
\usepackage{hyperref}

\newcommand{\pip}{$\pi^{+}$}
\newcommand{\pim}{$\pi^{-}$}
\newcommand{\kap}{K$^{+}$}
\newcommand{\kam}{K$^{-}$}
\newcommand{\pbar}{$\rm\overline{p}$}

\newcommand{\s}{$\sqrt{s}$}
\newcommand{\pt}{\ensuremath{p_{\rm t}}}
\newcommand{\dedx}{d$E$/d$x$}

\newcommand{\pp}{pp}

\usepackage{preprintcover}

\PreprintIdNumber{CERN-PH-EP-2010-085}
\PreprintCoverPaperTitle{Production of pions, kaons and protons in \pp\ collisions\\
 at \s ~= 900~GeV
with ALICE at the LHC}
\PreprintCoverAbstract{The production of \pip, \pim, \kap, \kam, p, and \pbar\
at mid-rapidity has been measured in proton-proton collisions at
\s~ = 900~GeV with the ALICE detector. Particle identification is
performed using the specific energy loss in the inner tracking
silicon detector and the time projection chamber.  In addition,
time-of-flight information is used to identify hadrons at higher
momenta. Finally, the distinctive kink topology of the weak decay
of charged kaons is used for an alternative measurement of the
kaon transverse momentum (\pt) spectra. Since these various
particle identification tools give the best separation
capabilities over different momentum ranges, the results are
combined to extract spectra from $\pt = 100$ MeV/$c$ to 2.5
GeV/$c$.  The measured spectra are further compared with
QCD-inspired models which yield a poor description. The total
yields and the mean \pt\ are compared with previous measurements,
and the trends as a function of collision energy are
discussed.}
\PreprintJournalName{EPJC}

\begin{document}
\hugehead

%
\title{Production of pions, kaons and protons in \pp\ collisions\\
 at \s ~= 900~GeV
with ALICE at the LHC }
\subtitle{ALICE collaboration}
\author{  
K.~Aamodt\inst{77} \and  
N.~Abel\inst{43} \and  
U.~Abeysekara\inst{75} \and  
A.~Abrahantes~Quintana\inst{42} \and  
A.~Abramyan\inst{112} \and  
D.~Adamov\'{a}\inst{85} \and  
M.M.~Aggarwal\inst{25} \and  
G.~Aglieri~Rinella\inst{40} \and  
A.G.~Agocs\inst{18} \and  
S.~Aguilar~Salazar\inst{63}
% ...
\renewcommand{\notesname}{Affiliation notes}  
\endnotetext[1]{Also at{ Laboratoire de Physique Corpusculaire (LPC), Clermont Universit\'{e}, Universit\'{e} Blaise Pascal, CNRS--IN2P3, Clermont-Ferrand, France}}  
\endnotetext[2]{Also at{ Frankfurt Institute for Advanced Studies, Johann Wolfgang Goethe-Universit\"{a}t Frankfurt, Frankfurt, Germany}}  
\endnotetext[3]{Now at{ Sezione INFN, Padova, Italy}}  
\endnotetext[4]{Now at{ Research Division and ExtreMe Matter Institute EMMI, GSI Helmholtzzentrum f\"{u}r Schwerionenforschung, Darmstadt, Germany}}  
\endnotetext[5]{Now at{ Institut f\"{u}r Kernphysik, Johann Wolfgang Goethe-Universit\"{a}t Frankfurt, Frankfurt, Germany}}  
\endnotetext[6]{Now at{ Physics Department, University of Cape Town, iThemba Laboratories, Cape Town, South Africa}}  
\endnotetext[7]{Now at{ National Institute for Physics and Nuclear Engineering, Bucharest, Romania}}  
\endnotetext[8]{Also at{ University of Houston, Houston, TX, United States}}  
\endnotetext[9]{Now at{ Faculty of Science, P.J.~\v{S}af\'{a}rik University, Ko\v{s}ice, Slovakia}}  
\endnotetext[10]{Now at{ European Organization for Nuclear Research (CERN), Geneva, Switzerland}}  
\endnotetext[11]{Now at{ Helsinki Institute of Physics (HIP) and University of Jyv\"{a}skyl\"{a}, Jyv\"{a}skyl\"{a}, Finland}}  
\endnotetext[12]{Now at{ Institut Pluridisciplinaire Hubert Curien (IPHC), Universit\'{e} de Strasbourg, CNRS-IN2P3, Strasbourg, France}}  
\endnotetext[13]{Now at{ Sezione INFN, Bari, Italy}}  
\endnotetext[14]{Now at{ Institut f\"{u}r Kernphysik, Westf\"{a}lische Wilhelms-Universit\"{a}t M\"{u}nster, M\"{u}nster, Germany}}  
\endnotetext[15]{Now at: University of Technology and Austrian Academy of Sciences, Vienna, Austria}  
\endnotetext[16]{Also at{ Lawrence Livermore National Laboratory, Livermore, CA, United States}}  
\endnotetext[17]{Also at{ European Organization for Nuclear Research (CERN), Geneva, Switzerland}}  
\endnotetext[18]{Now at { Secci\'{o}n F\'{\i}sica, Departamento de Ciencias, Pontificia Universidad Cat\'{o}lica del Per\'{u}, Lima, Peru}}  
\endnotetext[19]{Deceased}  
\endnotetext[20]{Now at{ Yale University, New Haven, CT, United States}}  
\endnotetext[21]{Now at{ University of Tsukuba, Tsukuba, Japan}}  
\endnotetext[22]{Also at { Centro Fermi -- Centro Studi e Ricerche e Museo Storico della Fisica ``Enrico Fermi'', Rome, Italy}}  
\endnotetext[23]{Now at{ Dipartimento Interateneo di Fisica `M.~Merlin' and Sezione INFN, Bari, Italy}}  
\endnotetext[24]{Also at{ Laboratoire de Physique Subatomique et de Cosmologie (LPSC), Universit\'{e} Joseph Fourier, CNRS-IN2P3, Institut Polytechnique de Grenoble, Grenoble, France}}  
\endnotetext[25]{Now at{ Dipartimento di Fisica Sperimentale dell'Universit\`{a} and Sezione INFN, Turin, Italy}}  
\endnotetext[26]{Now at{ Physics Department, Creighton University, Omaha, NE, United States}}  
\endnotetext[27]{Now at{ Commissariat \`{a} l'Energie Atomique, IRFU, Saclay, France}}  
\endnotetext[28]{Also at{ Department of Physics, University of Oslo, Oslo, Norway}}  
\endnotetext[29]{Now at{ Physikalisches Institut, Ruprecht-Karls-Universit\"{a}t Heidelberg, Heidelberg, Germany}}  
\endnotetext[30]{Now at{ Institut f\"{u}r Kernphysik, Technische Universit\"{a}t Darmstadt, Darmstadt, Germany}}  
\endnotetext[31]{Now at{ Department of Physics and Technology, University of Bergen, Bergen, Norway}}  
\endnotetext[32]{Now at{ Physics Department, University of Athens, Athens, Greece}}  
\endnotetext[33]{Also at{ Institut f\"{u}r Kernphysik, Westf\"{a}lische Wilhelms-Universit\"{a}t M\"{u}nster, M\"{u}nster, Germany}}  
\endnotetext[34]{Now at{ SUBATECH, Ecole des Mines de Nantes, Universit\'{e} de Nantes, CNRS-IN2P3, Nantes, France}}  
\endnotetext[35]{Now at{ Universit\'{e} de Lyon, Universit\'{e} Lyon 1, CNRS/IN2P3, IPN-Lyon, Villeurbanne, France}}  
\endnotetext[36]{Now at: Centre de Calcul IN2P3, Lyon, France}  
\endnotetext[37]{Now at{ Variable Energy Cyclotron Centre, Kolkata, India}}  
\endnotetext[38]{Also at{ Dipartimento di Fisica dell'Universit\`{a} and Sezione INFN, Padova, Italy}}  
\endnotetext[39]{Also at{ Sezione INFN, Bologna, Italy}}  
\endnotetext[40]{Also at Dipartimento di Fisica dell\'{ }Universit\`{a}, Udine, Italy}  
\endnotetext[41]{Also at Wroc{\l}aw University, Wroc{\l}aw, Poland} 
\endnotetext[42]{Now at Dipartimento di Fisica dell'Universit\`{a} and Sezione INFN, Padova, Italy}
\bigskip  
\theendnotes  
\section*{Collaboration institutes}  
}  

\institute{
Dipartimento di Scienze e Tecnologie Avanzate dell'Universit\`{a} del Piemonte Orientale and Gruppo Collegato INFN, Alessandria, Italy
\and
Department of Physics Aligarh Muslim University, Aligarh, India
\and
Nikhef, National Institute for Subatomic Physics, Amsterdam, Netherlands
\and
Physics Department, University of Athens, Athens, Greece
\and
Dipartimento Interateneo di Fisica `M.~Merlin' and Sezione INFN, Bari, Italy
\and
Sezione INFN, Bari, Italy
\and
China Institute of Atomic Energy, Beijing, China
\and
Department of Physics and Technology, University of Bergen, Bergen, Norway
\and
Faculty of Engineering, Bergen University College, Bergen, Norway
\and
Lawrence Berkeley National Laboratory, Berkeley, CA, United States
\and
Institute of Physics, Bhubaneswar, India
\and
School of Physics and Astronomy, University of Birmingham, Birmingham, United Kingdom
\and
Dipartimento di Fisica dell'Universit\`{a} and Sezione INFN, Bologna, Italy
\and
Sezione INFN, Bologna, Italy
\and
Faculty of Mathematics, Physics and Informatics, Comenius University, Bratislava, Slovakia
\and
Institute of Space Sciences (ISS), Bucharest, Romania
\and
National Institute for Physics and Nuclear Engineering, Bucharest, Romania
\and
KFKI Research Institute for Particle and Nuclear Physics, Hungarian Academy of Sciences, Budapest, Hungary
\and
Dipartimento di Fisica dell'Universit\`{a} and Sezione INFN, Cagliari, Italy
\and
Sezione INFN, Cagliari, Italy
\and
Universidade Estadual de Campinas (UNICAMP), Campinas, Brazil
\and
Physics Department, University of Cape Town, iThemba Laboratories, Cape Town, South Africa
\and
Dipartimento di Fisica e Astronomia dell'Universit\`{a} and Sezione INFN, Catania, Italy
\and
Sezione INFN, Catania, Italy
\and
Physics Department, Panjab University, Chandigarh, India
\and
Laboratoire de Physique Corpusculaire (LPC), Clermont Universit\'{e}, Universit\'{e} Blaise Pascal, CNRS--IN2P3, Clermont-Ferrand, France
\and
Department of Physics, Ohio State University, Columbus, OH, United States
\and
Niels Bohr Institute, University of Copenhagen, Copenhagen, Denmark
\and
The Henryk Niewodniczanski Institute of Nuclear Physics, Polish Academy of Sciences, Cracow, Poland
\and
Universidad Aut\'{o}noma de Sinaloa, Culiac\'{a}n, Mexico
\and
Research Division and ExtreMe Matter Institute EMMI, GSI Helmholtzzentrum f\"{u}r Schwerionenforschung, Darmstadt, Germany
\and
Institut f\"{u}r Kernphysik, Technische Universit\"{a}t Darmstadt, Darmstadt, Germany
\and
Wayne State University, Detroit, MI, United States
\and
Joint Institute for Nuclear Research (JINR), Dubna, Russia
\and
Frankfurt Institute for Advanced Studies, Johann Wolfgang Goethe-Universit\"{a}t Frankfurt, Frankfurt, Germany
\and
Institut f\"{u}r Kernphysik, Johann Wolfgang Goethe-Universit\"{a}t Frankfurt, Frankfurt, Germany
\and
Laboratori Nazionali di Frascati, INFN, Frascati, Italy
\and
Gangneung-Wonju National University, Gangneung, South Korea
\and
Petersburg Nuclear Physics Institute, Gatchina, Russia
\and
European Organization for Nuclear Research (CERN), Geneva, Switzerland
\and
Laboratoire de Physique Subatomique et de Cosmologie (LPSC), Universit\'{e} Joseph Fourier, CNRS-IN2P3, Institut Polytechnique de Grenoble, Grenoble, France
\and
Centro de Aplicaciones Tecnol\'{o}gicas y Desarrollo Nuclear (CEADEN), Havana, Cuba
\and
Kirchhoff-Institut f\"{u}r Physik, Ruprecht-Karls-Universit\"{a}t Heidelberg, Heidelberg, Germany
\and
Physikalisches Institut, Ruprecht-Karls-Universit\"{a}t Heidelberg, Heidelberg, Germany
\and
Hiroshima University, Hiroshima, Japan
\and
University of Houston, Houston, TX, United States
\and
Physics Department, University of Rajasthan, Jaipur, India
\and
Physics Department, University of Jammu, Jammu, India
\and
Helsinki Institute of Physics (HIP) and University of Jyv\"{a}skyl\"{a}, Jyv\"{a}skyl\"{a}, Finland
\and
Bogolyubov Institute for Theoretical Physics, Kiev, Ukraine
\and
University of Tennessee, Knoxville, TN, United States
\and
Saha Institute of Nuclear Physics, Kolkata, India
\and
Variable Energy Cyclotron Centre, Kolkata, India
\and
Fachhochschule K\"{o}ln, K\"{o}ln, Germany
\and
Faculty of Science, P.J.~\v{S}af\'{a}rik University, Ko\v{s}ice, Slovakia
\and
Institute of Experimental Physics, Slovak Academy of Sciences, Ko\v{s}ice, Slovakia
\and
Laboratori Nazionali di Legnaro, INFN, Legnaro, Italy
\and
Secci\'{o}n F\'{\i}sica, Departamento de Ciencias, Pontificia Universidad Cat\'{o}lica del Per\'{u}, Lima, Peru
\and
Lawrence Livermore National Laboratory, Livermore, CA, United States
\and
Division of Experimental High Energy Physics, University of Lund, Lund, Sweden
\and
Centro de Investigaciones Energ\'{e}ticas Medioambientales y Tecnol\'{o}gicas (CIEMAT), Madrid, Spain
\and
Instituto de Ciencias Nucleares, Universidad Nacional Aut\'{o}noma de M\'{e}xico, Mexico City, Mexico
\and
Instituto de F\'{\i}sica, Universidad Nacional Aut\'{o}noma de M\'{e}xico, Mexico City, Mexico
\and
Centro de Investigaci\'{o}n y de Estudios Avanzados (CINVESTAV), Mexico City and M\'{e}rida, Mexico
\and
Institute for Nuclear Research, Academy of Sciences, Moscow, Russia
\and
Institute for Theoretical and Experimental Physics, Moscow, Russia
\and
Moscow Engineering Physics Institute, Moscow, Russia
\and
Russian Research Centre Kurchatov Institute, Moscow, Russia
\and
Indian Institute of Technology, Mumbai, India
\and
Institut f\"{u}r Kernphysik, Westf\"{a}lische Wilhelms-Universit\"{a}t M\"{u}nster, M\"{u}nster, Germany
\and
SUBATECH, Ecole des Mines de Nantes, Universit\'{e} de Nantes, CNRS-IN2P3, Nantes, France
\and
Yale University, New Haven, CT, United States
\and
Budker Institute for Nuclear Physics, Novosibirsk, Russia
\and
Oak Ridge National Laboratory, Oak Ridge, TN, United States
\and
Physics Department, Creighton University, Omaha, NE, United States
\and
Institut de Physique Nucl\'{e}aire d'Orsay (IPNO), Universit\'{e} Paris-Sud, CNRS-IN2P3, Orsay, France
\and
Department of Physics, University of Oslo, Oslo, Norway
\and
Dipartimento di Fisica dell'Universit\`{a} and Sezione INFN, Padova, Italy
\and
Sezione INFN, Padova, Italy
\and
Faculty of Nuclear Sciences and Physical Engineering, Czech Technical University in Prague, Prague, Czech Republic
\and
Institute of Physics, Academy of Sciences of the Czech Republic, Prague, Czech Republic
\and
Institute for High Energy Physics, Protvino, Russia
\and
Benem\'{e}rita Universidad Aut\'{o}noma de Puebla, Puebla, Mexico
\and
Pusan National University, Pusan, South Korea
\and
Nuclear Physics Institute, Academy of Sciences of the Czech Republic, \v{R}e\v{z} u Prahy, Czech Republic
\and
Dipartimento di Fisica dell'Universit\`{a} `La Sapienza' and Sezione INFN, Rome, Italy
\and
Sezione INFN, Rome, Italy
\and
Commissariat \`{a} l'Energie Atomique, IRFU, Saclay, France
\and
Dipartimento di Fisica `E.R.~Caianiello' dell'Universit\`{a} and Sezione INFN, Salerno, Italy
\and
California Polytechnic State University, San Luis Obispo, CA, United States
\and
Departamento de F\'{\i}sica de Part\'{\i}culas and IGFAE, Universidad de Santiago de Compostela, Santiago de Compostela, Spain
\and
Universidade de S\~{a}o Paulo (USP), S\~{a}o Paulo, Brazil
\and
Russian Federal Nuclear Center (VNIIEF), Sarov, Russia
\and
Department of Physics, Sejong University, Seoul, South Korea
\and
Yonsei University, Seoul, South Korea
\and
Technical University of Split FESB, Split, Croatia
\and
V.~Fock Institute for Physics, St. Petersburg State University, St. Petersburg, Russia
\and
Institut Pluridisciplinaire Hubert Curien (IPHC), Universit\'{e} de Strasbourg, CNRS-IN2P3, Strasbourg, France
\and
University of Tokyo, Tokyo, Japan
\and
Dipartimento di Fisica dell'Universit\`{a} and Sezione INFN, Trieste, Italy
\and
Sezione INFN, Trieste, Italy
\and
University of Tsukuba, Tsukuba, Japan
\and
Dipartimento di Fisica Sperimentale dell'Universit\`{a} and Sezione INFN, Turin, Italy
\and
Sezione INFN, Turin, Italy
\and
Nikhef and Institute for Subatomic Physics of Utrecht University, Utrecht, Netherlands
\and
Universit\'{e} de Lyon, Universit\'{e} Lyon 1, CNRS/IN2P3, IPN-Lyon, Villeurbanne, France
\and
Soltan Institute for Nuclear Studies, Warsaw, Poland
\and
Warsaw University of Technology, Warsaw, Poland
\and
Purdue University, West Lafayette, IN, United States
\and
Zentrum f\"{u}r Technologietransfer und Telekommunikation (ZTT), Fachhochschule Worms, Worms, Germany
\and
Hua-Zhong Normal University, Wuhan, China
\and
Yerevan Physics Institute, Yerevan, Armenia
\and
Rudjer Bo\v{s}kovi\'{c} Institute, Zagreb, Croatia
}

\date{Received: \today / Revised version: date}
%
\abstract{The production of \pip, \pim, \kap, \kam, p, and \pbar\
at mid-rapidity has been measured in proton-proton collisions at
\s~ = 900~GeV with the ALICE detector. Particle identification is
performed using the specific energy loss in the inner tracking
silicon detector and the time projection chamber.  In addition,
time-of-flight information is used to identify hadrons at higher
momenta. Finally, the distinctive kink topology of the weak decay
of charged kaons is used for an alternative measurement of the
kaon transverse momentum (\pt) spectra. Since these various
particle identification tools give the best separation
capabilities over different momentum ranges, the results are
combined to extract spectra from $\pt = 100$ MeV/$c$ to 2.5
GeV/$c$.  The measured spectra are further compared with
QCD-inspired models which yield a poor description. The total
yields and the mean \pt\ are compared with previous measurements,
and the trends as a function of collision energy are
discussed.  } 
\PACS{{25.75.Dw}{Particle and resonance production}
     \and
     {13.85.Ni}{Inclusive production with identified hadrons}
}   

\maketitle
%

\section{Introduction}
\label{intro}

In \pp\ collisions at ultra-relativistic energies the bulk of the
particles produced at mid-rapidity have transverse momenta, \pt,
below 1 GeV/$c$. Their production is not  calculable from first
principles via perturbative Quantum Chromodynamics, and is not
well modelled at lower collision
energies. 
This low \pt\ particle production, and species composition, must
therefore be measured, providing crucial  input for the modelling
of hadronic interactions and the hadronization process. It is
important to study the bulk production of particles as a function
of both \pt\ and particle species. With the advent of \pp\
collisions at the Large Hadron Collider (LHC)  at CERN a new
energy regime is being explored, where particle production from
hard interactions which  are predominantly gluonic in nature, is
expected to play an increasing role. Such data will provide extra
constraints on the modelling of fragmentation functions. The data
will also serve as a reference for the heavy-ion measurements.

The ALICE detector~\cite{Alessandro:2006yt,Carminati:2004fp} is
designed to perform measurements in the high-multiplicity
environment expected in central lead-lead collisions at
$\sqrt{s_{\rm NN}}$ = 5.5 TeV at the LHC and to identify particles
over a wide range of momenta. As such, it is ideally suited to
perform these measurements also in pp collisions.

This paper presents the transverse momentum spectra and yields of
identified particles at mid-rapidity from the first \pp\ collisions
collected in the autumn of 2009, during the commissioning of the
LHC,  at \s~= 900 GeV. The evolution of particle production in
\pp\ collisions with collision energy is studied by comparing
to data from previous experiments.

We report \pip, \pim, \kap, \kam, p, and \pbar\ distributions,
identified via several independent techniques utilizing specific
energy loss, \dedx, information from the Inner Tracking System
(ITS) and the Time Projection Chamber (TPC), and velocity
measurements in the Time-Of-Flight array (TOF). The combination of
these methods provides particle identification over the transverse
momentum range $0.1~\mathrm{GeV}/c <$ \pt $< 2.5~\mathrm{GeV}/c$.
Charged kaons, identified via kink topology of their weak decays
in the TPC, provide a complementary measurement over a similar
\pt\ range. All reported particle yields are for primary
particles, namely those directly produced in the collision
including the products of strong and electromagnetic decays but
excluding weak decays of strange particles.

The paper is organized as follows: In Section 2, the ALICE
detectors relevant for these studies, the experimental conditions,
and the corresponding analysis techniques are described. Details
of the event and particle selection are presented. In Section 3,
the \pip, \pim, \kap, \kam, p, and \pbar\ inclusive spectra and
yields, obtained by combining the various techniques described in
Section 2, are presented. The results are compared with
calculations from QCD-inspired models and the \pt-dependence of
ratios of particle yields, e.g.~K/$\pi$ and p/$\pi$, are
discussed. Comparisons with data from other experiments at
different \s\ are made and the evolution of the ratio of strange
to non-strange hadrons with collision energy is discussed.
Finally, in Section 4 the results are summarized.

\section{Experimental setup and data analysis}
\label{exp_data}

\subsection{The ALICE detector}
\label{experiment}

The ALICE detector and its expected performance are described in
detail in \cite{Alessandro:2006yt,Carminati:2004fp,ALICE-JINST}.
For the analyses described in this paper the following detectors
are used: the ITS, the TPC and the TOF detector. These detectors
are positioned in a solenoidal magnetic field of $B$~=~0.5~T and
have a common pseudo-rapidity coverage of $-0.9 < \eta < 0.9$. Two
forward scintillator hodoscopes (VZERO) are used for triggering
purposes. They are placed on either side of the interaction
region, covering regions $2.8 < \eta < 5.1$ and $-3.7 < \eta <
-1.7$.

\subsubsection{The Inner Tracking System}

The ITS is the closest of the central barrel detectors to the beam
axis. It is composed of six cylindrical layers of silicon
detectors. The two innermost layers are equipped with pixel
detectors (SPD), followed by two layers of drift detectors (SDD)
and two layers of double-sided silicon strip detectors (SSD). The
innermost layer is at 3.9 cm from the beam axis, while the outer
layer is at 43.0 cm.

The ITS provides high-resolution space points that allow the
extension of tracks reconstructed in the TPC towards the
interaction vertex, thus improving momentum and angular
resolution. The four layers equipped with SDD and SSD also provide
a measurement of the specific energy loss \dedx . The SPD yields
an on-line measure of the multiplicity by counting the number of
chips that have one or more hits (fast-OR), which is included in
the minimum-bias trigger
logic~\cite{ALICE-JINST,AglieriRinella:2007hm}. The ITS is also
used as a stand-alone tracker to reconstruct charged particles
with momenta below 200 MeV/$c$ that are deflected or decay before
reaching the TPC, and to recover tracks crossing dead regions of
the TPC.
A detailed description of the three sub-systems can be found
in~\cite{ALICE-JINST}. The \dedx\ measurement in the SDD and SSD
has been calibrated using cosmic ray data and pp events
\cite{Alessandro:2010rq}. The 2198 ITS modules have been aligned
using survey information, cosmic-ray tracks and pp data with the
methods described in~\cite{:2010ys}. The fraction of active
modules per layer in the present setup is around 80\% in the SPD
and 90\% - 95\% both in SDD
and SSD. 

\subsubsection{The Time Projection Chamber}

The TPC is  the main tracking device. It is a large volume, high
granularity, cylindrical detector with an outer radius of 2.78 m
and a length of 5.1 m. The active volume extends from 0.85 m to
2.47 m in radius. It covers 2$\pi$ in azimuth and $|\eta|<0.9$ in
polar angle for the full radial track length. Accepting  one third
of the full radial track length extends the range  to $|\eta|<$
1.5. The 90 m$^3$ drift volume is  filled with  a Ne (85.7\%),
CO$_2$ (9.5\%), and N$_2$ (4.8\%) gas mixture. A high voltage
central membrane splits the drift region in two halves, resulting
in a maximal drift time of 94 $\mu$s. Each of the two read-out planes is composed of 18 inner
and 18 outer chambers with a total of 159 pad rows, resulting in a
total of  557 568 pads which are read out separately. The position
resolution in $r\phi$ direction varies from 1100 $\mu$m to 800
$\mu$m when going from the inner to the outer radius. Along the
beam axis ($z$, also the drift direction) the resolution ranges between 1250 $\mu$m and 1100
$\mu$m. A maximum of 159 clusters can be measured along a track in
the TPC. For a detailed description see~\cite{Alme:2010ke}.

\subsubsection{The Time-Of-Flight Detector }
The TOF detector consists of 18 azimuthal sectors, each containing
91 Multi-gap Resistive Plate Chambers (MRPCs) distributed in five
gas-tight modules. It is positioned at 370-399 cm from the beam
axis. The region $\rm{260^\circ}$ $< \phi $ $< \rm{320^{\circ}}$
at $\eta \sim 0$ is not covered in order to minimize the material
in front of the Photon Spectrometer, which is not used in this
analysis. The MRPC detectors are installed with a projective
geometry along the beam direction, minimizing the variation of the
flight path of particles across the sensitive area of the
detector. Each MRPC is segmented into 96 read-out pads (2.5
$\times$ 3.5 $\rm{cm^2}$ size), resulting in a total of 152928
channels. Test beam results demonstrated that the intrinsic time
resolution of the detector is better than 50~$\rm{ps}$, dominated
by electronic effects and the time resolution of the
time-to-digital converters~\cite{Akindinov:2009zz}. Results from
the TOF commissioning with cosmic rays are described in references
~\cite{Akindinov:2006hs,Akindinov:2010zza,Akindinov:2010zz}. In the present setup,
9.6\% of the readout channels were inactive due to failures in the
high- or low-voltage systems or in the readout electronics. The
fraction of noisy channels, identified during data taking by
online monitoring and excluded from the subsequent reconstruction,
was below 0.1\%.

\subsection{Event selection and normalization}
\label{events}

The data presented in this paper were collected during the
commissioning of the LHC at CERN in the autumn of 2009, with \pp\
collisions at $\sqrt{s}=900$~GeV. The collider was run with four
bunches per beam, resulting in two bunch crossings per beam
circulation period (89~$\mu$s) at the ALICE interaction point. The
remaining two bunches per beam were not collided at ALICE, and
served to estimate the contribution of beam-gas interactions. The
average event rate was a few Hz, so the fraction of pile-up events
was negligible.

The analysis is based on a sample of $ \sim300$k inelastic \pp\
collisions.  The online trigger selection requires a signal in
either of the VZERO counters or at least one hit in either of the
SPD layers. The selection was improved offline with recomputed
trigger input quantities using the time average over all VZERO
hits and a suppression of noisy channels. The contamination from
beam-induced background is rejected offline using the timing
information of the VZERO and by cutting on the correlation between
the number of clusters and track segments (tracklets) in the SPD
detector \cite{Aamodt:2010ft,Aamodt:2010my}. Selected events are
further required to contain a reconstructed primary vertex. The
vertex reconstruction efficiency calculated via Monte-Carlo
simulations is 96.5\% for events with one reconstructed track and
approaches unity for events with more than two tracks.

The results presented in this paper are normalized to inelastic
\pp\ collisions, employing the strategy described in
\cite{Aamodt:2010ft,Aamodt:2010my}.  In order to reduce the
extrapolation and thus the systematic uncertainty on the
normalization, the sample of selected events used for
normalization includes triggered events without reconstructed
tracks or vertices. Those ev\-ents still contain a small
contamination from very low multiplicity beam-induced background
or accidentals from the trigger, which are not rejected by the
selections described above.  This contamination is of the order of
4\% and is subtracted using the control triggers.  From the
analysis of empty bunch events the random contribution from cosmic
rays is found to be negligible. The number of selected events is
then converted to the number of inelastic collisions after
correcting for the trigger efficiency, which is determined from
the Monte-Carlo simulation, scaling the cross section for
diffractive processes to the measurements of
UA5~\cite{Ansorge:1986xq}. The subtraction of beam-gas events and
the efficiency correction partially compensate each other: the
overall correction factor is about 5\% with a systematic
uncertainty of about 2\%, coming mainly from the uncertainties in
the modelling of diffraction in the event
generators. 

In order to compare to previous experimental results, which are
only published for the~ non-single-diffractive (NSD) class, in
Section~\ref{results}, we scale our spectra for the measured ratio
$\left. {\rm d}N_{ch}/{\rm d}\eta\right|_{NSD} / \left. {\rm
d}N_{ch}/{\rm
    d}\eta\right|_{INEL} \simeq 1.185$~\cite{Aamodt:2010ft}. PYTHIA and PHOJET
simulations indicate that the \pt-dependence of the ratio of spectra  for
NSD and inelastic collisions is less than 5\% in the reported range.
Particle ratios are found
to be insensitive to the conversion from inelastic to
non-single-diffractive events.

\subsection{Track selection}
\label{tracks}

The identified particle spectra were measured independently with
the ITS, TPC and TOF, and combined in the final stage of the
analysis. The rapidity range $|y|<0.5$ was used for all analyses
except for the kink analysis ($|y|<0.7$).

For the TPC and TOF analyses, tracks reconstructed in the TPC are
used. The TPC has full acceptance for tracks with $|\eta|<0.9$.
However, shorter tracks at higher $\eta$ can still be used for
physics analysis, in particular protons with a transverse momentum
of \pt = 400~MeV/$c$ and $|y| = 0.5$ which correspond to
$|\eta|=1.1$. To ensure high tracking efficiency and
\dedx-resolution, while keeping the contamination from secondaries
and fakes low, tracks are required to have at least 80 clusters,
and a $\chi^{2}$ of the momentum fit that is smaller than 4 per
cluster. Since each cluster in the TPC provides two degrees of
freedom and the number of parameters of the track fit is much
smaller than the number of clusters, the $\chi^2$ cut is
approximately 2 per degree of freedom. In addition, at least two
clusters in the ITS must be associated to the track, out of which
at least one is from the SPD. Tracks are further rejected based on
their distance-of-closest approach (DCA) to the reconstructed
event vertex. The cut is implemented as a function of \pt\ to
correspond to about seven (five) standard deviations in the
transverse (longitudinal) coordinate, taking into account the
\pt-dependence of the impact parameter resolution. These selection
criteria are tuned to select primary charged particles with high
efficiency while minimizing the contributions from weak decays,
conversions and secondary hadro\-nic interactions in the detector
material. The DCA resolution in the data is found to be in good
agreement with the Monte-Carlo simulations that are used for
efficiency corrections (see next Section).

Tracks reconstructed in the TPC are extrapolated to the sensitive
layer of the TOF and a corresponding signal is searched for. The
channel with the center closest to the track extrapolation point
is selected if the distance is less than 10 cm. This rather weak
criterion results in a high matching efficiency while keeping the
fraction of wrongly associated tracks below 1\% in the low-density
environment presented by \pp\ collisions.

The \dedx\ measurements in the ITS are used to identify hadrons in
two independent analyses, based on different tracking algorithms.  One
analysis uses the ITS-TPC combined tracking, while the other is
based on ITS stand-alone tracks. The combined ITS-TPC tracking result serves as a cross-check of both the
ITS stand-alone and the TPC results in the overlap region. The ITS
stand-alone analysis extends the acceptance to lower \pt\ than the
TPC or ITS-TPC analyses.

The combined ITS-TPC analysis uses the same track selection
criteria as the TPC only analysis, with the additional requirement
of at least four clusters in the ITS, out of which at least one
must be in the SPD and at least three in SSD+SDD. This further
reduces the contamination of secondaries and provides high
resolution on track impact parameter and optimal resolution on the
\dedx.
The ITS stand-alone tracking uses a similar selection, with a
different $\chi^2$ selection and a different DCA selection. In the
current tracking algorithm, ITS clusters are assigned a larger
 position error to account for residual misalignment of the detector.
As a result, the $\chi^2$ values are not properly normalized, but
the selection was adjusted to be equivalent to the TPC $\chi^2$
selection by inspecting the distributions. The DCA cut in the ITS
analysis uses the same \pt-dependent parametrization as for TPC
tracks, but with different parameters to account for the different
resolution.

\subsection{Monte-Carlo Calculations}

The efficiency and other correction factors including acceptance
(jointly called \textit{efficiency} in the following discussion)
used in this paper are calculated from a Monte-Carlo simulation,
based on over two million events produced with the PYTHIA 6.4
event generator~\cite{Sjostrand:2006za} (tune D6T~\cite{D6T}),
propagated through the detector with the GE\-ANT3~\cite{:1994zzo}
transport code. Dead and noisy channels as well as beam position
and spread have been taken into account.  A simulation based on
the PHOJET event generator~\cite{Engel:1995sb} is also used as a
cross check.

GEANT3 is known to reproduce the absorption cross sections of
hadrons incorrectly. The transport code FLU\-KA contains a more
accurate description of these cross
sections~\cite{Bendiscioli:1994uv,
Zhang:1996ev,Klempt:2002ap}, and a dedicated simulation is used to
calculate a correction to the GEANT3 efficiency
calculation~\cite{Aamodt:2010dx}. This is relevant mainly for
antiprotons at low \pt, where the correction is on the order of
10\%. For other particles and at higher \pt, the difference
between GEANT and FLUKA calculations is negligible.

\subsection{Particle Identification}
\label{sec:part-ident}

The \dedx{} and TOF signals are used for particle identification
as a function of the momentum $p$, whereas the final spectra are
given as a function of the transverse momentum \pt.

In the case of the TPC and ITS analyses, particles were identified via
the specific energy loss \dedx. Unique identification on a
track-by-track basis is possible in regions of momentum where the
bands are clearly separated from each other. In overlapping areas,
particle identification is still possible on a statistical basis using
fits to the energy loss distribution in each \pt-bin. The fits are
performed on the distribution of the difference between the measured
and the expected energy deposition for tracks within the selected
rapidity range $|y|$ $<$ 0.5. 
This compensates for
the very steep slope of the Bethe-Bloch in the $1/\beta^2$ region
which would make the \dedx-distribution in a simple \pt\ or $p$-slice
non-Gaussian. The calculated expected energy loss depends on the
measured track momentum $p$ and the assumed mass for the particle. The
procedure is therefore repeated three times for the entire set of
tracks, assuming the pion, kaon, and proton mass.

In the TPC analysis, the difference \newline [\dedx]$_{\rm meas} -
$[\dedx$(pid, p_{tot})]_{\rm calc}$ is used. For the ITS the
difference of the logarithm of the measured and calculated energy
deposit $\ln[$\dedx$_{\rm
meas}]-\ln[$\dedx($pid$,$p_{tot}\rm)_{\rm calc}]$ is taken to
suppress the non-gaussian tails originating from the smaller
number of \dedx\ measurements.

In the case of the TOF, the identification is based on the
time-of-flight information. The procedure for the extraction of
the raw yields differs slightly from the one used for TPC and ITS,
and is described in Section~\ref{sec:TOF}.

\subsubsection{Particle identification in the ITS}
\label{ITS}

In both the ITS stand-alone and in the ITS-TPC analyses, the
\dedx\ measurement from the SDD and the SSD is used to identify
particles. The stand-alone tracking result extends the momentum
range to lower \pt{} than can be measured in the TPC, while the
combined tracking provides a better momentum resolution.

The energy loss measurement in each layer of the ITS is corrected
for the track length in the sensitive volume using tracking
information. In the case of SDD clusters, a linear correction for
the dependence of the reconstructed raw charge as a function of
drift time due to the combined effect of charge diffusion and zero
suppression is also applied~\cite{Alessandro:2010rq}. For each
track, \dedx\ is calculated using a truncated mean: the average of
the lowest two points in case four points are measured, or a
weighted sum of the lowest (weight 1) and the second lowest point
(weight 1/2), in case only three points are measured.

\begin{figure}[h]
\resizebox{0.48\textwidth}{!}{%
  \includegraphics{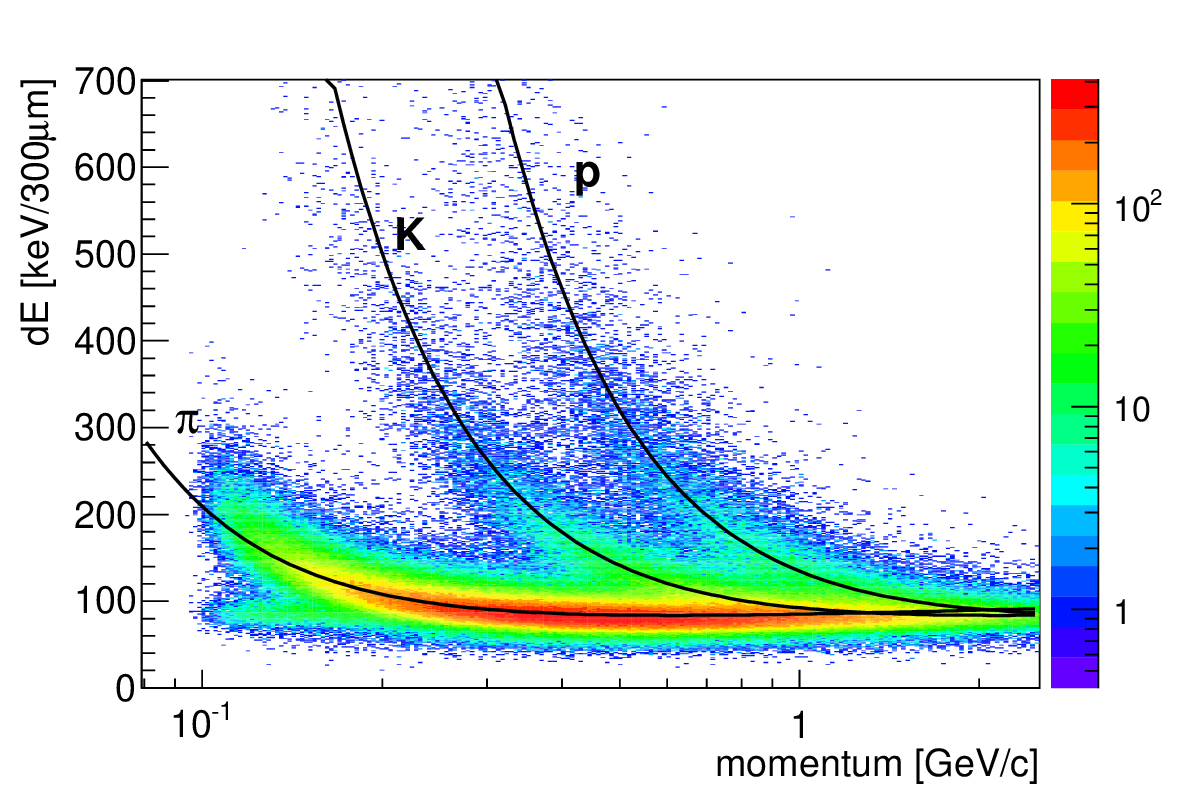}
} \caption{(Color online) Specific energy loss \dedx\ vs.~momentum
for tracks measured with the ITS. The solid lines are a
parametrization (from \cite{Back:2006tt}) of the detector response
based on the Bethe-Bloch formula.} \label{fig:ITSdedx}
\end{figure}

Figure~\ref{fig:ITSdedx} shows the truncated mean \dedx\ for the
sample of ITS stand-alone tracks along with the PHOBOS
parametrization of the most probable value~\cite{Back:2006tt}.

\begin{figure*}[ht!]
\centering
\resizebox{0.85\textwidth}{!}{%
\includegraphics{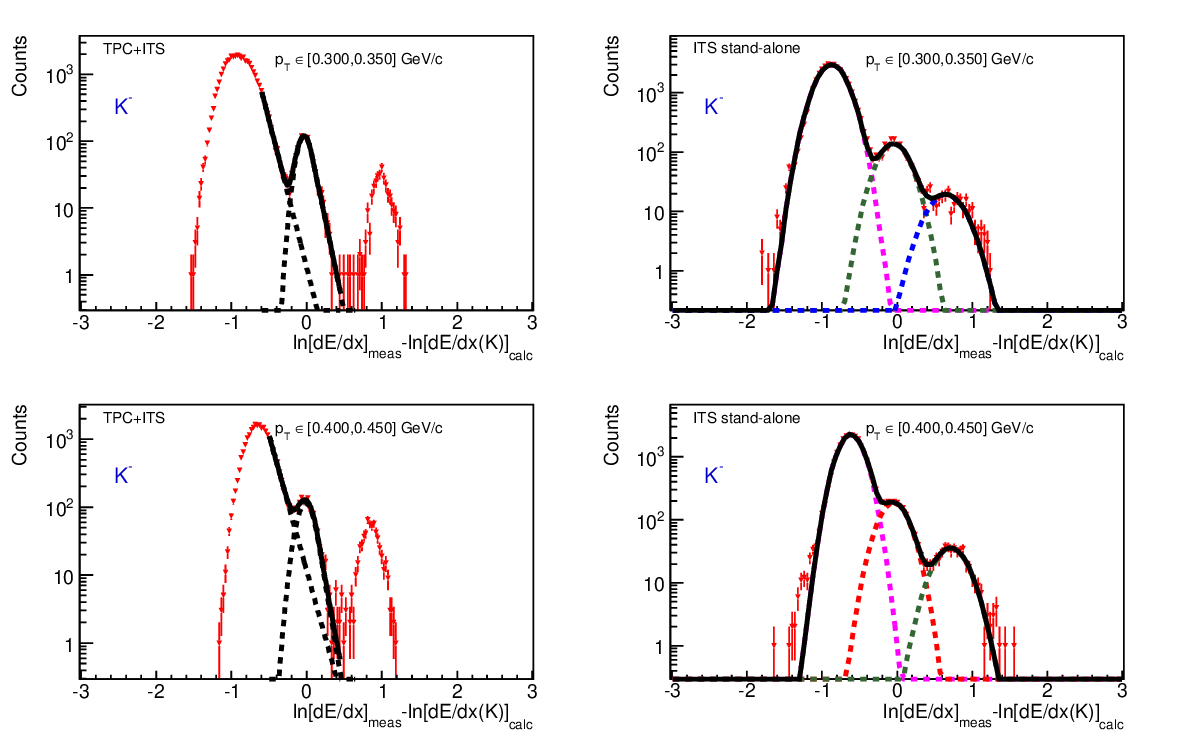}
} \caption{(Color online) Distribution of ln[\dedx]$_{\rm
meas}-$ln[\dedx({\rm K})]$_{\rm calc}$ measured with the ITS in
the two \pt-ranges, 300--350 MeV/$c$ (upper panels) and 400-450
MeV/$c$ (lower panels), using the kaon mass hypothesis. The left
panels show the result for ITS-TPC combined tracks, while the
right panels show the ITS stand-alone result.  The lines indicate
fits as described in the text.} \label{its:3gausITS}
\end{figure*}

For the ITS stand-alone track sample, the histograms are fitted
with three Gaussians and the integral of the Gaussian centered at
zero is used as the raw yield of the corresponding hadron species.
In a first step, the peak widths $\sigma$ of the peaks are extracted as a
function of \pt\ for pions and protons in the region where their
\dedx\ distributions do not overlap with the kaon (and
electron) distribution. For kaons, the same procedure is used at low \pt,
where they are well separated. The \pt-dependence of the peak width is then
extrapolated to higher \pt\ with the same functional form used to
describe the pions and protons. The resulting
parametrizations of the \pt\ dependence of $\sigma$ are
used to constrain the fits of the ln[\dedx]
distributions to extract the raw yields.

For the ITS-TPC combined track sample, a non-Gau-ssian tail is
visible. This tail is a remnant of the tail of the Landau
distribution for energy loss. It was verified using simulations
that the shape and size of the tail are compatible with the
expectations for a truncated mean using two out of four samples.
The tail is not as pronounced for the ITS stand-alone track
sample, due to the limited momentum resolution. The distribution
is fitted with a combination of a Gaussian and an exponential
function for the main peak and another exponential function to
describe the tail of a background peak. This functional form
provides an accurate description of the peak shape in the detector
simulation, as well as the measured shape.

Examples of \dedx\
distributions are shown in Fig.~\ref{its:3gausITS} for negative
tracks using the kaon mass hypothesis in two different \pt\
intervals for both ITS stand-alone tracks (right panels) and
ITS-TPC combined tracks (left panels).

\paragraph{Efficiency correction}

The raw hadron yields extracted from the fits to the \dedx\
distributions are corrected for the reconstruction efficiency
determined from Monte-Carlo simulations, applying the same
analysis criteria to the simulated events  as to the data.
Secondary particles from interactions in the detector material and
strange particle decays have been subtracted from the yield of
both simulated and real data. The fraction of secondaries after
applying the track impact-parameter cut depends on the hadron
species and amounts
 to 1-3\% for pions and 5-10\% for protons depending on \pt. The
secondary-to-primary ratio has been estimated by fitting the
measured track impact-parameter distributions with three
components, prompt particles, secondaries from strange particle
decays and secondaries produced in the detector material for each
hadron species. Alternatively, the contamination from secondaries
have been determined using Monte-Carlo samples, after rescaling
the $\Lambda$ yield to the measured values~\cite{strange}. The
difference between these two procedures is about 3\% for protons
and is negligible for other particles.

\begin{figure}[h]
\resizebox{0.45\textwidth}{!}{%
\includegraphics{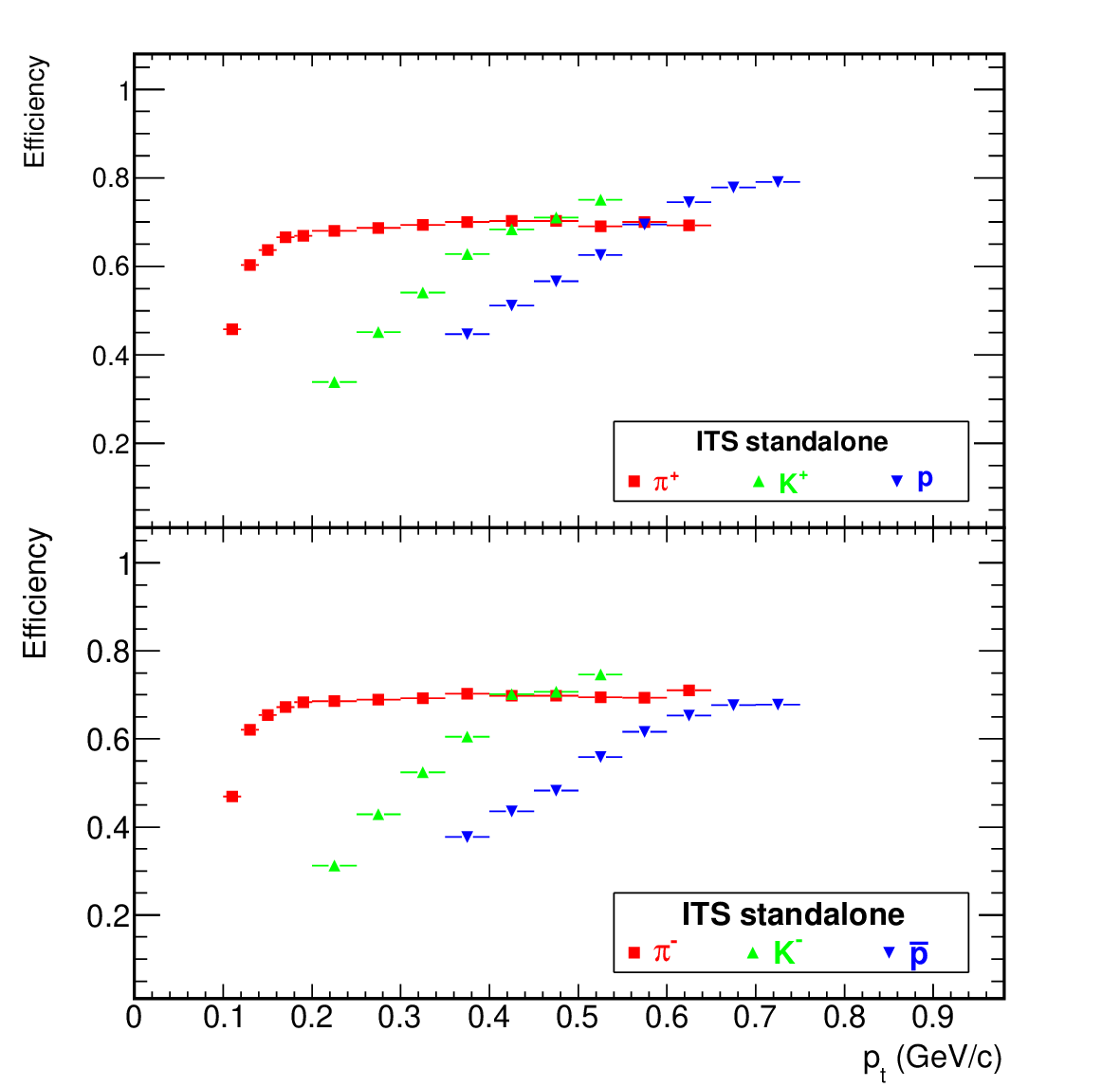}
} \caption{(Color online) Efficiency for pions, kaons and protons
for the ITS stand-alone analysis as obtained from Monte-Carlo
simulations. } \label{its:eff}
\end{figure}

Figure~\ref{its:eff} shows the total reconstruction efficiency for
 primary tracks in the ITS stand-alone, including the effects of detector and tracking
efficiency, the track selection cuts and residual contamination in
the fitting procedure, as determined from the Monte-Carlo
simulation. This efficiency is used to correct the measured raw
yields after subtraction of the contributions from secondary
hadrons. The measured spectra are corrected for the efficiency of
the primary vertex reconstruction with the SPD using the ratio
between generated primary spectra in simulated events with a reconstructed
vertex and events passing the trigger conditions.

Systematic errors are summarized in Table~\ref{tab:itssyst}.  The
systematic uncertainty from secondary contamination has been
estimated by repeating the full analysis chain with different cuts
on the track impact parameter and by comparing the two alternative
estimates outlined above.  The effect of the uncertainty in the
material budget has been estimated by modifying the material
budget in the Monte-Carlo simulations by $\pm 7$\%, which is the
present uncertainty of the ITS material budget.  The systematic
contribution from the fitting procedure to the ln[\dedx]$_{\rm
meas} - $ln[\dedx(i)]$_{\rm calc}$ distributions has been
estimated by varying the fit condition and by comparing to an
independent analysis using a track-by-track identification
approach based on the distance between the measured and expected
\dedx\ values normalized to its resolution.  The residual
imperfections in the description of the ITS detector modules and
dead areas in the simulation introduce another uncertainty in the
ITS tracking efficiency. This is estimated by varying the cuts on
the number of clusters and on the track $\chi^2$ both in data and
in Monte-Carlo simulations.

In the lowest \pt-bins, a larger systematic error has been
assigned to account for the steep slope of the tracking efficiency
as a function of the particle transverse momentum (see
Fig.~\ref{its:eff}).

\begin{table}
\centering \caption{Summary of systematic errors in the efficiency
correction of the ITS analysis.} \label{tab:itssyst}
\begin{tabular}{lccc}
\hline\noalign{\smallskip}
systematic errors & $\pi^{\pm}$ & K$^{\pm}$ & p and \pbar \\
\noalign{\smallskip}\hline\noalign{\smallskip}
secondary contamination & negl. & negl.  & negl. \\
from material & & & \\
\noalign{\smallskip}\hline\noalign{\smallskip}
secondary contamination & $<1$\% & negl. & 3\% \\
from weak decay & & & \\
\noalign{\smallskip}\hline\noalign{\smallskip}
material budget & & &\\
highest \pt\ bin  & $<1$\% & $<1$\% & 1\% \\
lowest \pt\ bin  & 5\% & 2\% & 3\% \\
\noalign{\smallskip}\hline\noalign{\smallskip}
ITS efficiency \\
all \pt\ bins & 2\% & 2\% &  2\% \\
lowest \pt\ bin & 12\% & 13\% & 11\% \\
\noalign{\smallskip}\hline\noalign{\smallskip}
ln(\dedx) distr.  & 1\% & 5\% & 3.5\% \\
fitting procedure & & & \\
\noalign{\smallskip}\hline
\end{tabular}
\end{table}

\subsubsection{Particle identification in the TPC}
\label{TPC}

\begin{figure}[h]
\resizebox{0.499\textwidth}{!}{
\includegraphics{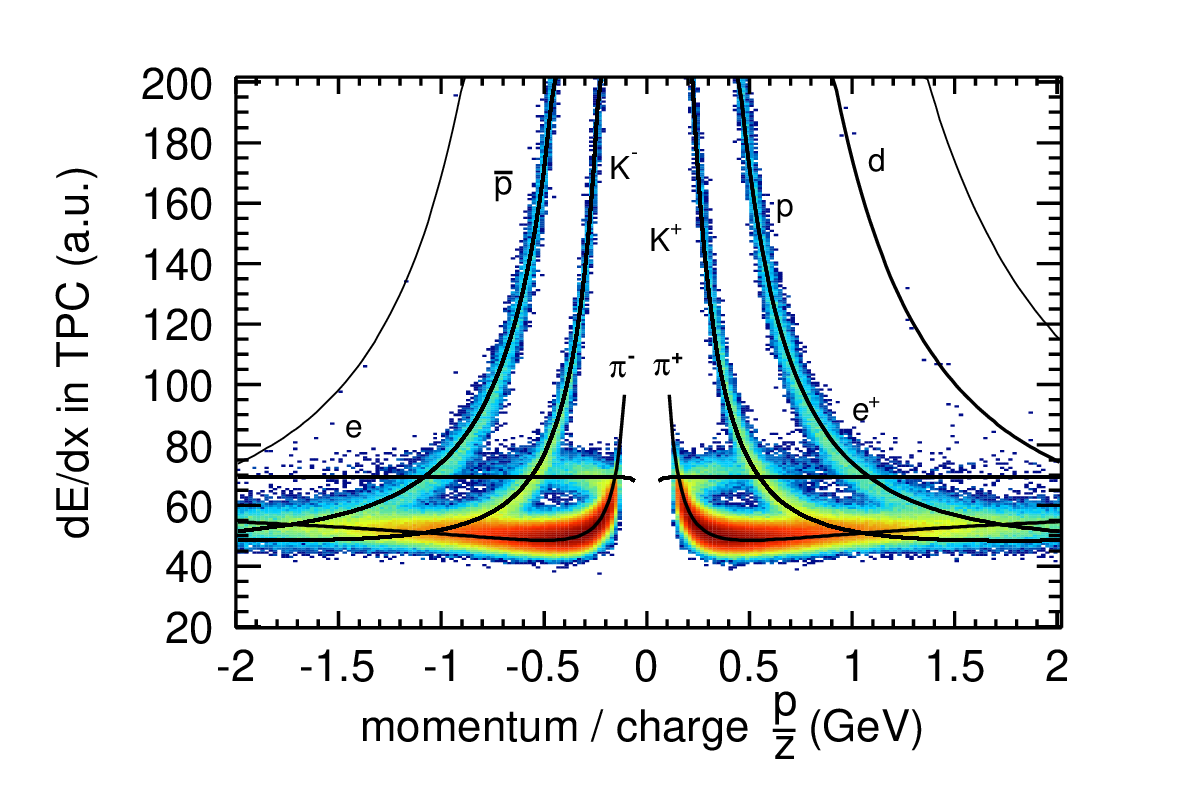}
}
\caption{(Color online) Specific energy loss \dedx\ vs.~momentum
for tracks measured with the ALICE TPC. The solid lines are a
parametrization of the Bethe-Bloch curve~\cite{ALEPH}. }
\label{tpc:dedx}
\end{figure}

Particle identification is based on the specific energy deposit of
each particle in the drift gas of the TPC, shown in
Fig.~\ref{tpc:dedx} as a function of momentum separately for
positive and negative charges.   The solid curves show the
calibration curves obtained by fitting the ALEPH parametrization
of the Bethe-Bloch curve~\cite{ALEPH} to the data points in
regions of clear separation.

The calibration parameters have mostly been determined and tested
via the analysis of cosmic rays. The pad-gain factors have been
measured using the decay of radioactive $^{83}_{36}$Kr gas
released into the TPC volume (for a detailed description
see~\cite{Alme:2010ke}).

As in the case of the ITS, a truncated-mean procedure is used to
determine \dedx\ (60\% of the points are kept). This reduces the
Landau tail of the \dedx{} distribution to the extent that it is
very close to a Gaussian distribution.

\begin{figure}[h]
\resizebox{0.47\textwidth}{!}{%
\includegraphics{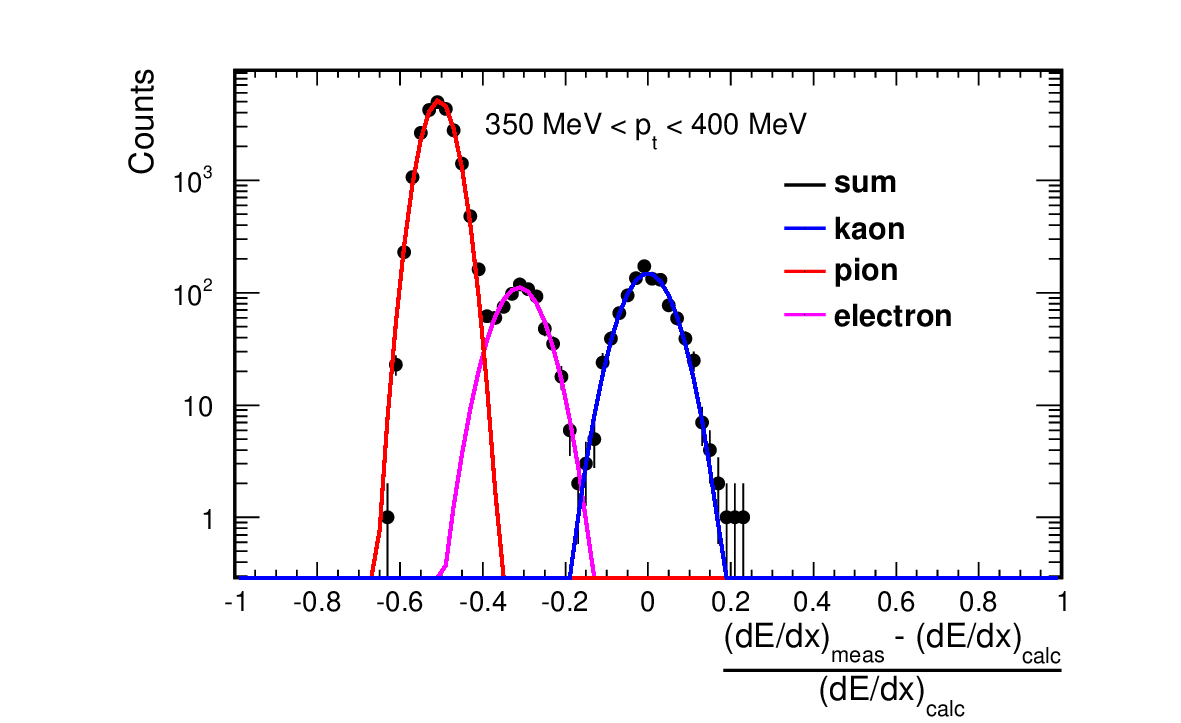}
}
\resizebox{0.47\textwidth}{!}{%
\includegraphics{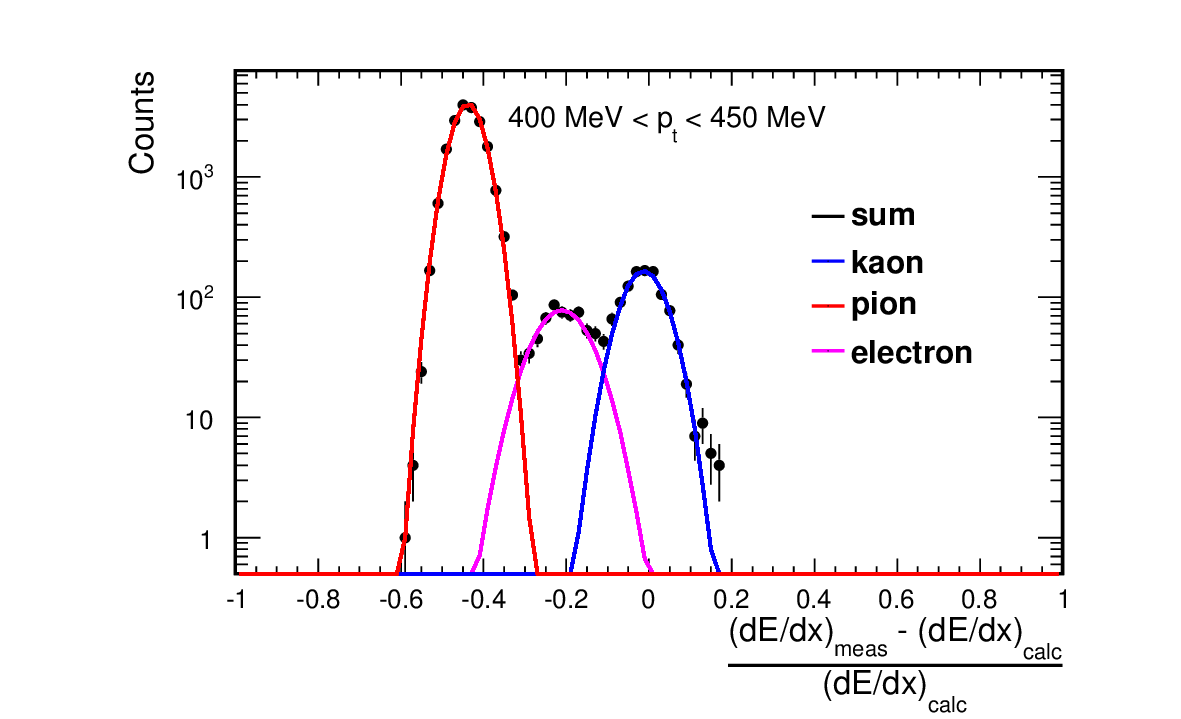}
}
\resizebox{0.47\textwidth}{!}{%
\includegraphics{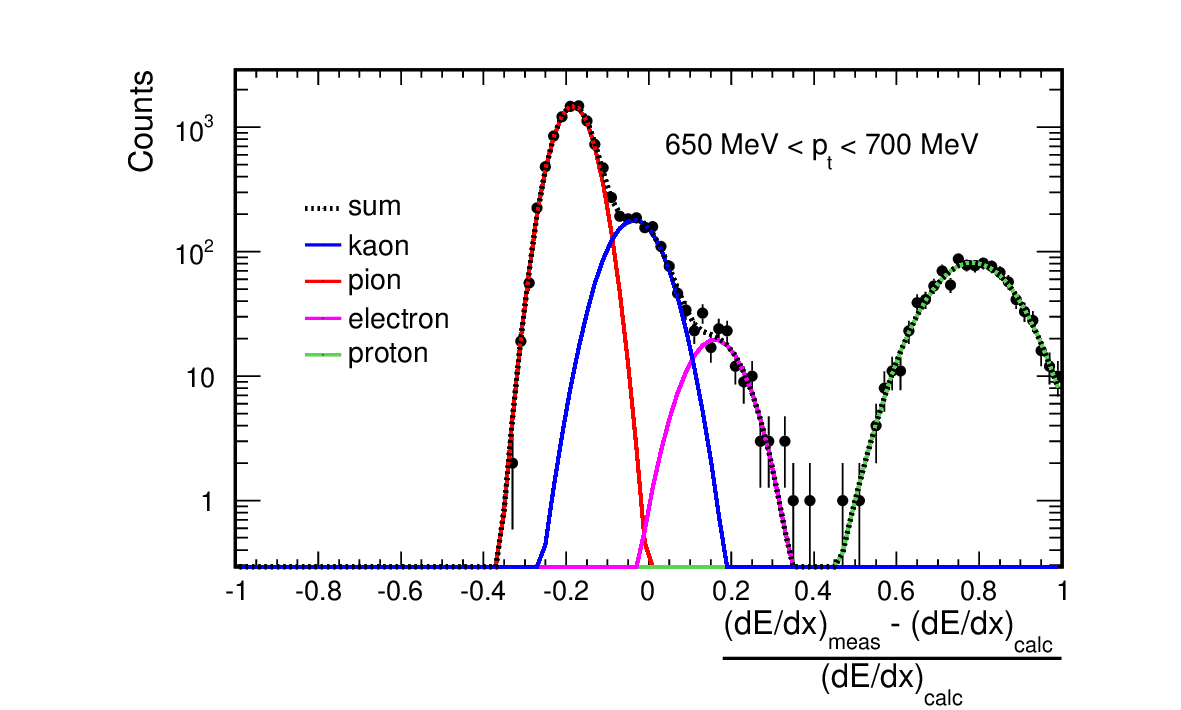}
} \caption{(Color online) Distribution of ([\dedx]$_{\rm
meas}-$[\dedx(kaon)]$_{\rm calc}$)/[\dedx(kaon)]$_{\rm calc}$
measured with the TPC for several \pt-bins showing the separation
power. The solid lines are Gaussian fits to the distributions. }
\label{tpc:pid}
\end{figure}

Examples of the \dedx\ distribution in some \pt\ bins are shown in
Fig.~\ref{tpc:pid}. The peak centered at zero is from kaons and
the other peaks are from other particle species. As the background
in all momentum bins is negligible, the integrals of the Gaussian
give the raw yields.

\paragraph{Efficiency correction}

\begin{figure}
\resizebox{0.45\textwidth}{!}{%
\includegraphics{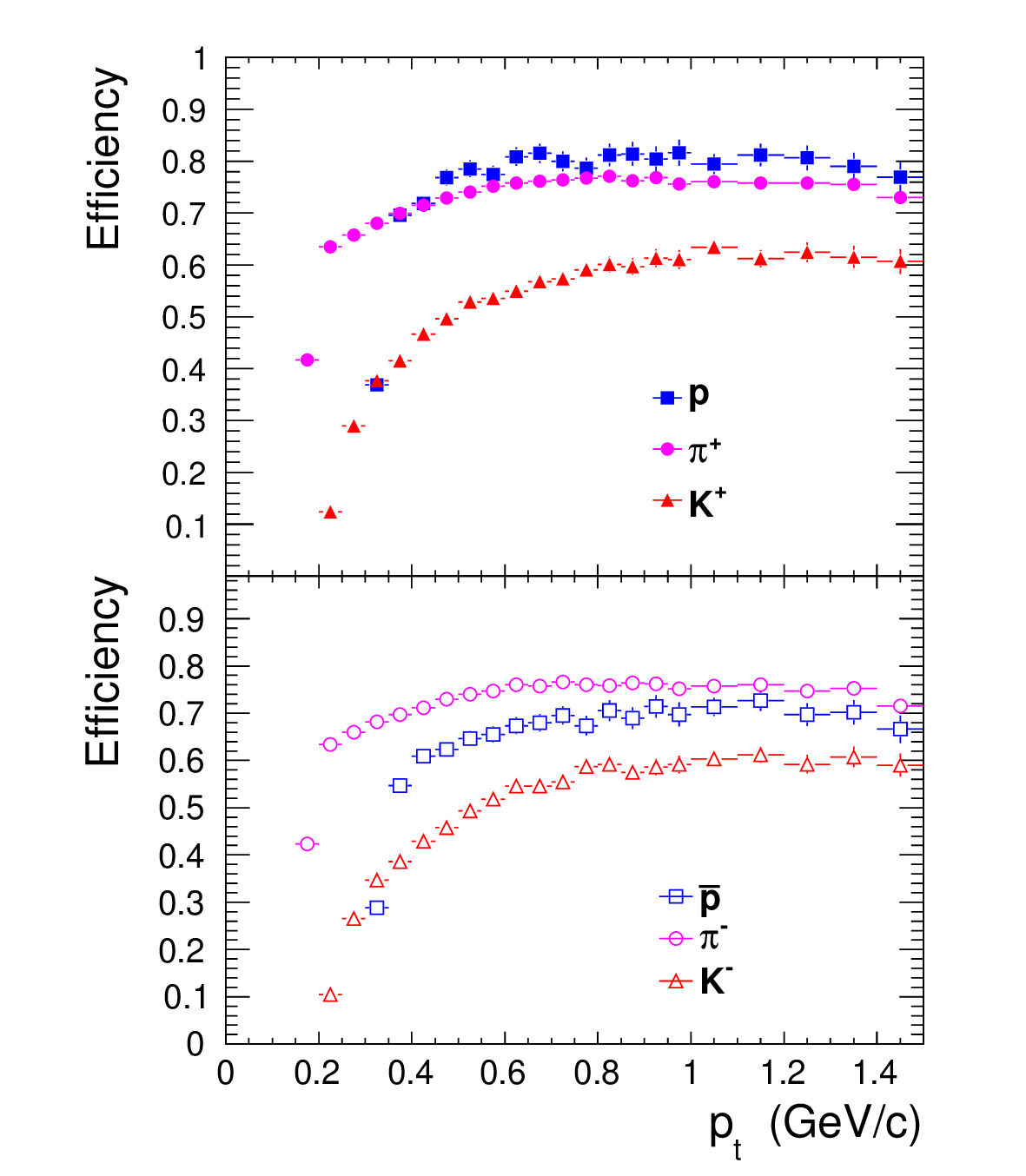}
} \caption{(Color online) Efficiency of charged pions, kaons, and
protons for the spectra extracted with the TPC.}
\label{tpc:pionEff}
\end{figure}

The raw hadron spectra are corrected for the reconstruction
efficiency, shown in Fig.~\ref{tpc:pionEff}, determined by doing
the same analysis on Monte-Carlo events. The efficiency is
calculated by comparing the number of reconstructed particles to
the number of charged primary particles from PYTHIA in the chosen
rapidity range.  For transverse momenta above 800 MeV/$c$ the
efficiency saturates at roughly 80\%. For kaons, the decay reduces
the efficiency by about 30\% at 250 MeV/$c$ and 12\% at 1.5
GeV/$c$. 
The range with a reconstruction efficiency lower than 60\% (for
pions and protons) is omitted for the analysis corresponding to a
low-\pt\ cut-off of 200 MeV/$c$ for pions, 250 MeV/$c$ for kaons,
and 400 MeV/$c$ for protons.

Protons are corrected for the contamination of secondaries from
material and of feed down from weak decays. The feed down was
determined by two independent methods. Firstly, the contamination
obtained from Monte-Carlo simulation was scaled such that it
corresponds to the measured yield of $\Lambda$s in the data
\cite{strange}. Secondly, the shape of the impact parameter
distribution was compared to the Monte-Carlo simulation. Weak
decays produce a non-Gaussian tail in the distribution of primary
particles whereas secondaries from material generate a flat
background \cite{Aamodt:2010dx}. The remaining difference between
the methods is included in the systematic error. The correction
for weak decays amounts to up to 14\% and the correction for
secondaries from material up to 4\% for protons with 400 MeV/$c <$
\pt\ $< 600$ MeV/$c$.  For other particle species and other
transverse momenta the contamination is negligible.

The systematic errors in the  track reconstruction and in the
removal of secondary particles have been estimated by varying the
number of standard deviations in the dis\-tan\-ce-to-vertex cut,
using a fixed cut of 3 cm instead of the variable one, and varying
the SPD-TPC matching cut. Their impact on the corrected spectra is
less than 5\%. The influence of the uncertainty in the material
budget has been examined by varying it by 7\%. This resulted in
the systematic errors given in Table~\ref{tab:tpcsyst}. The
uncertainty due to a possible deviation from a Gaussian shape has
been established by comparing the multi-Gauss fit with a
3-$\sigma$ band in well separated regions. The precision of the
kink rejection is estimated to be within 3\%.

The correction for the event selection bias has been tested with
two event generators, PYTHIA~\cite{Sjostrand:2006za,D6T} and
PHOJET~\cite{Engel:1995sb} and the corresponding uncertainty is
less than 1\%.

\begin{table}
\caption{Summary of systematic errors in the efficiency correction
in the TPC analysis.} \label{tab:tpcsyst}
\begin{tabular}{cccc}
\hline\noalign{\smallskip}
systematic errors & $\pi^{\pm}$ & K$^{\pm}$ & p and \pbar   \\
\noalign{\smallskip}\hline\noalign{\smallskip}

secondary contamination & negl. & negl.   & $<$ 2\% \\
from material & & & \\
\noalign{\smallskip}\hline\noalign{\smallskip}

secondary contamination & $<$ 4\% & - & $<$ 10\% \\
from weak decay & & & \\
\noalign{\smallskip}\hline\noalign{\smallskip}

energy loss and & $<$ 1\% & $<$ 1\% &  $<$ 2\% \\
absorption in material & & & \\
\noalign{\smallskip}\hline\noalign{\smallskip}

kink rejection & negl. & $<$ 3\% & - \\
\noalign{\smallskip}\hline\noalign{\smallskip}

non-Gaussianity of & negl. & negl. & negl. \\
dE/dx signal & & & \\
\noalign{\smallskip}\hline\noalign{\smallskip}


matching to ITS & & $<$ 3\% & \\
\noalign{\smallskip}\hline
\end{tabular}
\end{table}

\subsubsection{Particle identification with the TOF}
\label{sec:TOF}

Particles reaching the TOF system are identified by measuring
their momentum and velocity simultaneously.

\begin{figure}
\resizebox{0.485\textwidth}{!}{%
 \includegraphics{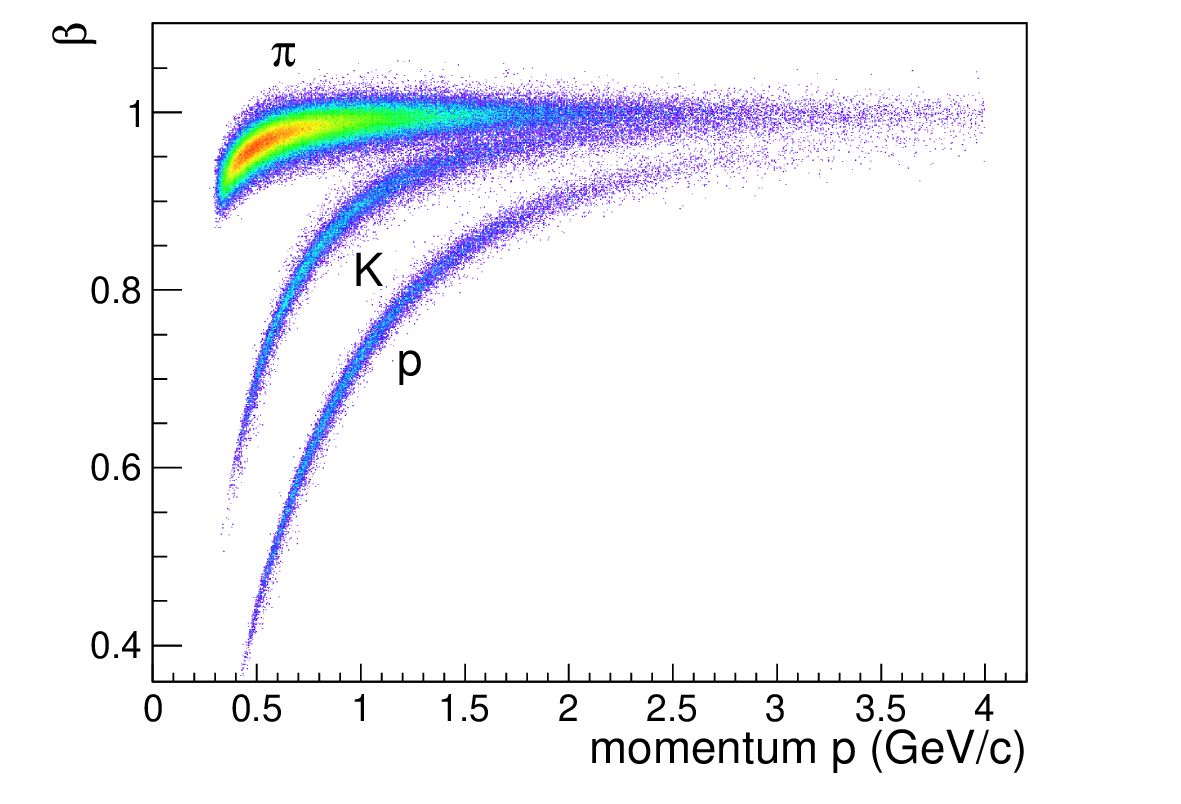}
} \caption{(Color online) $\beta$ of tracks of particles measured
by TOF vs. their momentum.} \label{TOF:beta_vs_p}
\end{figure}

The velocity $\beta= L/t_{\rm TOF}$ is obtained from the measured
time of flight $t_{\rm{TOF}}$ and the reconstructed flight path
$L$ along the track trajectory between the point of closest
approach to the event vertex and the TOF sensitive surface. The
measured velocities are shown as a function of the momentum $p$ at
the vertex in Fig.~\ref{TOF:beta_vs_p}. The bands corresponding to
charged pions, kaons and protons are clearly visible. The width of
the bands reflects the overall time-of-flight resolution of about
180 ps, which depends on the TOF timing signal resolution, the
accuracy of the reconstructed flight path and the uncertainty of
the event start time, $t_{0}^{ev}$. This last contribution is
related to the uncertainty in establishing the absolute time of
the collision. In the present sample this fluctuated with respect
to the nominal time signal from the LHC with a $\sigma$ of about
140 ps due to the finite size of the bunches.

To improve the overall time-of-flight resolution, the TOF
information itself is used to determine $t_{0}^{ev}$ in events
having at least three tracks with an associated TOF signal. This
is done with a combinatorial algorithm which compares the TOF
times with the calculated times of the tracks for each event for
different mass hypotheses. Using this procedure, the start-time
has been improved for 44\% of the tracks having an associated TOF
signal and is rather independent on the momentum of the tracks. In
this way the precision on the event start-time is about 85 ps on
average.

Finally, tracks whose particle identity as determined from the
TOF information is not compatible with the one inferred from the
\dedx\ signal in the TPC within five $\sigma$ have been removed.
This TOF-TPC compatibility criterion rejects about 0.6\% of the
tracks and further reduces the small contamination coming from
tracks incorrectly associated with a TOF signal.


\begin{figure}[ht]
\begin{center}
\resizebox{0.43\textwidth}{!}{%
\includegraphics{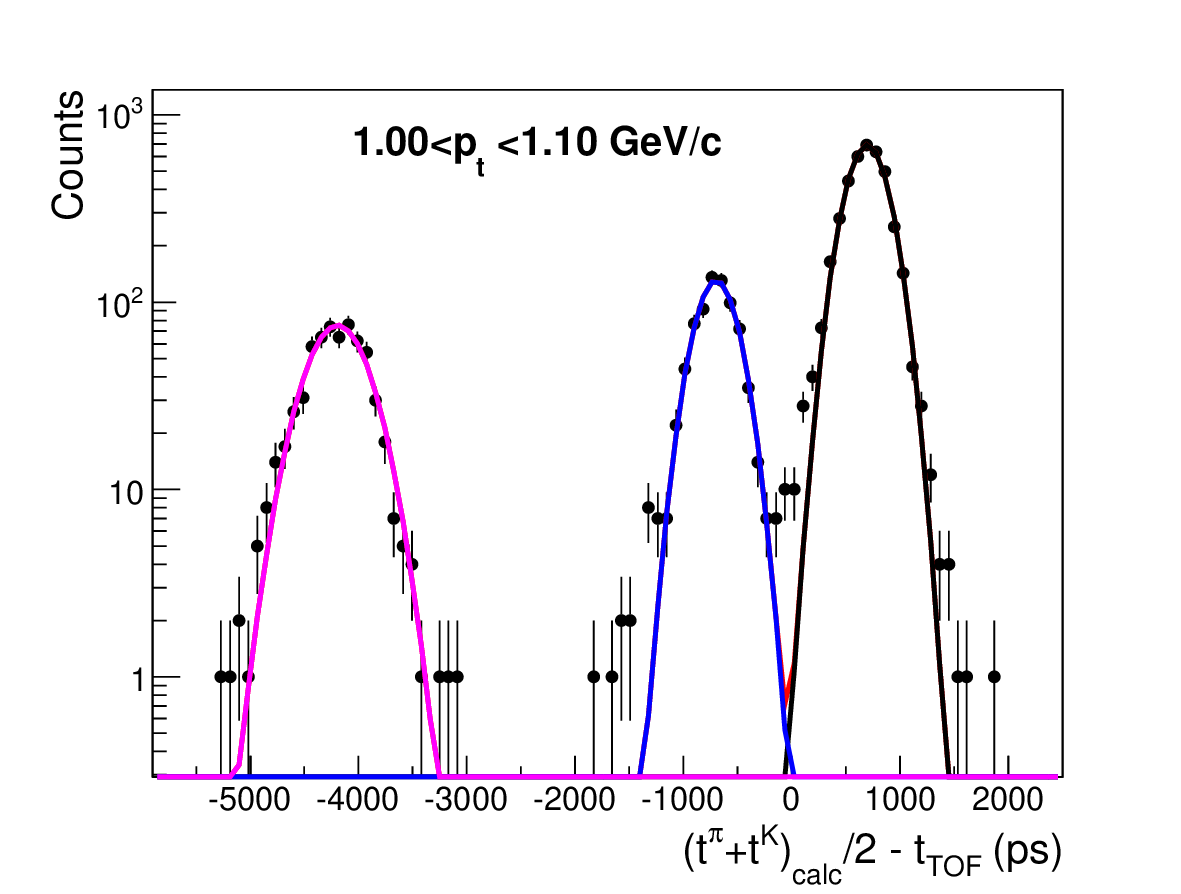}
}
\resizebox{0.43\textwidth}{!}{%
\includegraphics{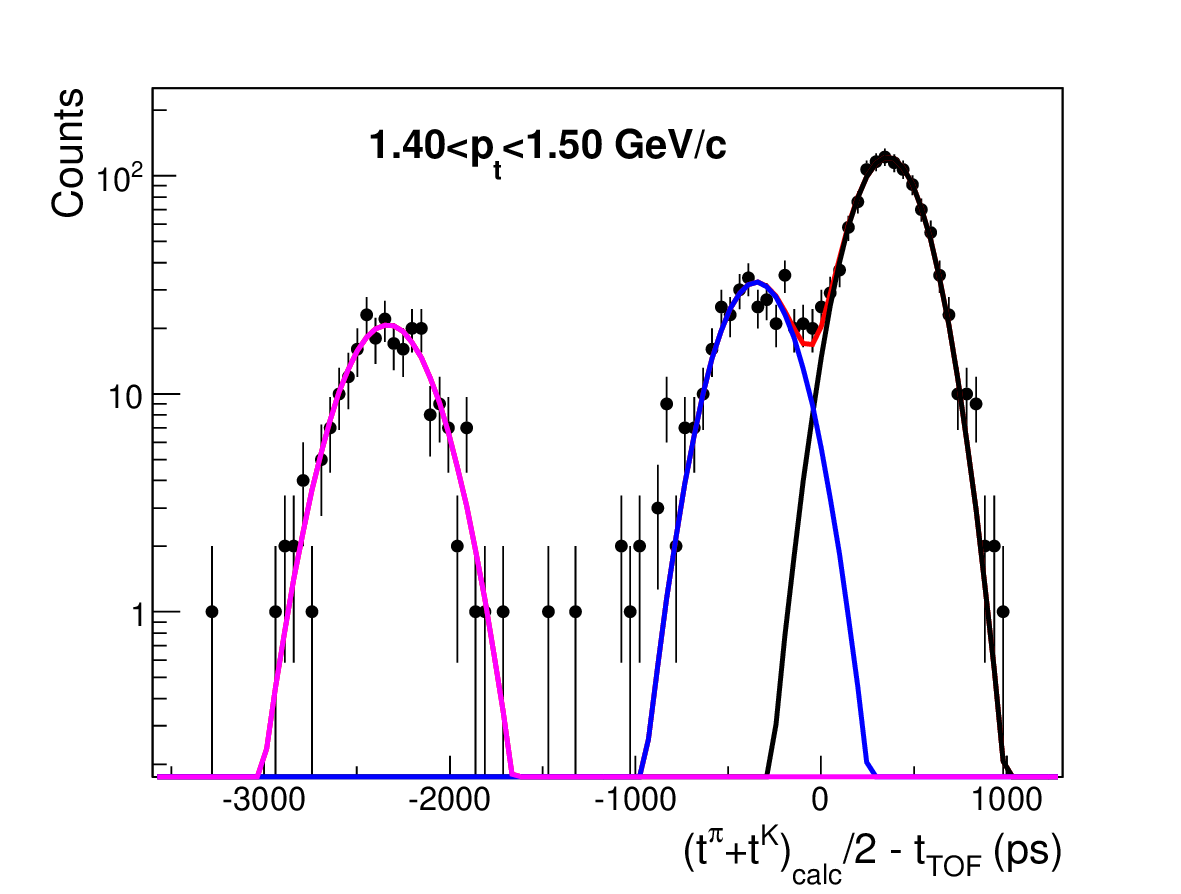}
}
\resizebox{0.43\textwidth}{!}{%
\includegraphics{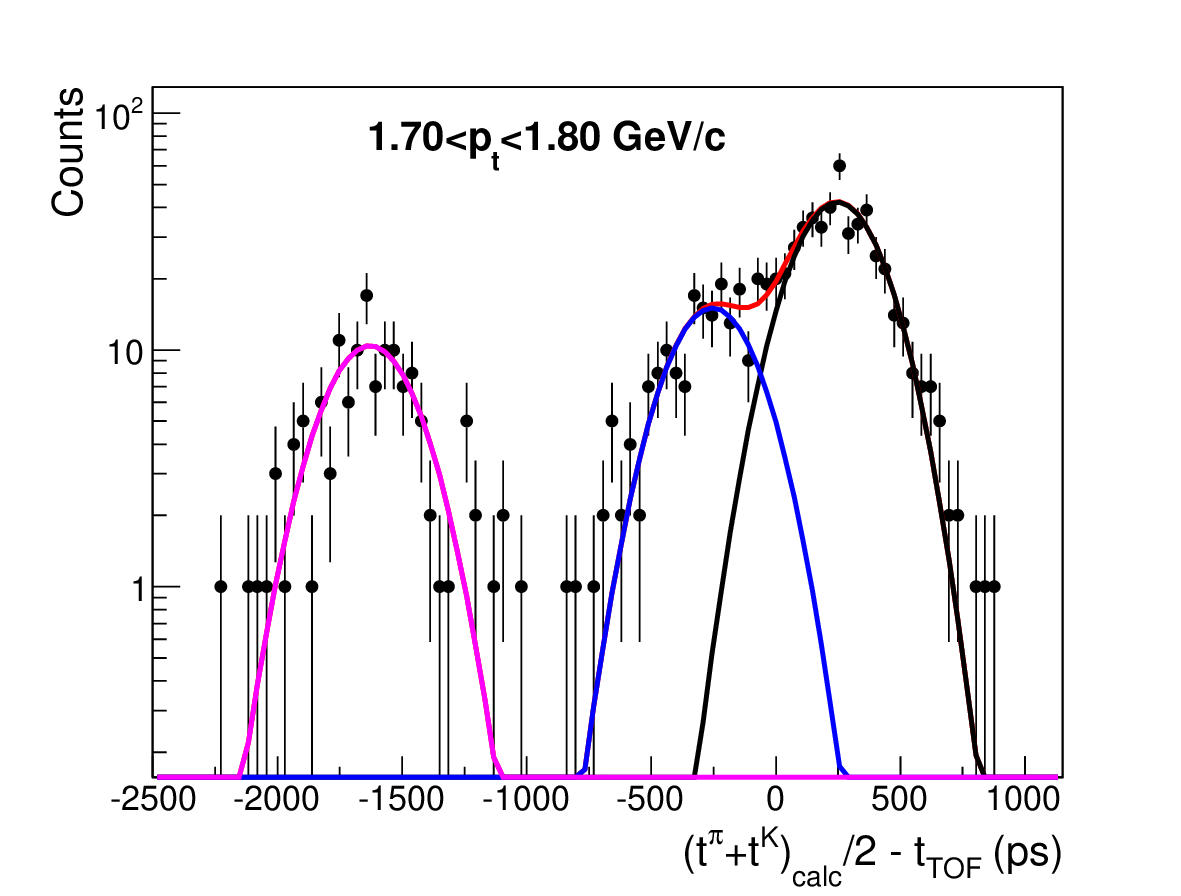}
} \caption{(Color online) Distribution of the time difference
between the measured TOF signal and the average  of the calculated
times for pions and kaons for several \pt-bins for positively
charged particles. The fits are performed using Gaussian shapes.}
\label{TOF:pid}
\end{center}
\end{figure}

For each particle species $i$, the expected time of flight $t_{\rm calc}^i$ is
calculated by summing up the
time-of-flight increments $\Delta t_k = \Delta l_k
\sqrt{p^2_k+m_i^2}/p_k$ at each tracking step, with $p_k$ being
the local value of the track momentum, $m_i$ the mass of the particle,
and $\Delta l_k$ the
track-length increment along its trajectory.
The yields of $\pi$, K and p are obtained from the simultaneous
fit of the distribution of the time difference $S$ between
measured $t_{\rm TOF}$ and the average between the calculated time
for pions and kaons
\begin{equation}
 S = (t^{\pi} + t^{\rm K})_{\rm calc}/2  - t_{\rm TOF}.
\end{equation}
The symmetric treatment of kaons and pions in the definition of
$S$ ensures that the kaon and pion peak are both Gaussian.
Extracting the yield for different species in a simultaneous fit
guarantees that the resulting number of pions, kaons and protons
matches the total number of tracks in the given momentum bin.

The distribution of the variable ${S}$ is shown in
Fig.~\ref{TOF:pid} for three different transverse momentum bins
for positive particles.
The curves show the results of the three-Gaussian fit used to
extract the raw yields. The integral of the fit result has been
constrained to the number of entries in the distribution, and the
means and the widths are allowed to vary within  5\% and 10\%,
respectively, of their nominal values. The only free parameters in
the fit are therefore the relative normalizations between the
Gaussians.

The raw yields are extracted in different \pt-bins using a
rapidity selection $|y_{\rm p}| < 0.5$, where $y_{\rm p}$ is the
rapidity calculated with the proton mass. For pions and kaons,
this condition results in a larger $y$-acceptance and in both
cases, the fraction outside of $|y| < 0.5$ has been subtracted in
each \pt-bin taking into account the $y$-distribution of the
yields within the pions and kaons peaks.

\paragraph{Efficiency correction}

\begin{figure}
\resizebox{0.485\textwidth}{!}{%
  \includegraphics{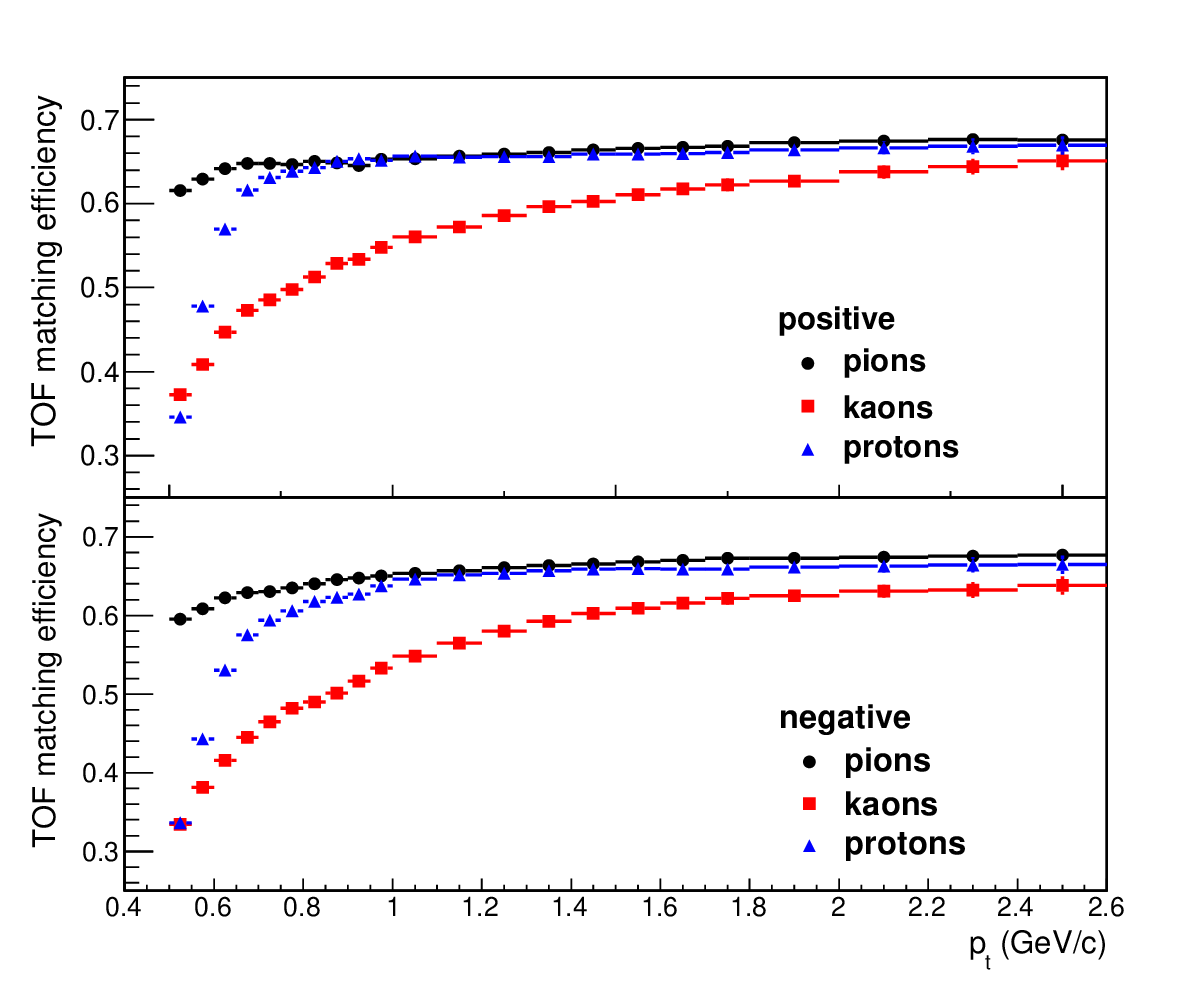} }
  \caption{(Color online) The TOF matching efficiency is shown for the three
particles, separately, for (top) positive and (bottom) negative
particles.} \label{TOF:matchingEff}
\end{figure}

Since the track selection used in the TOF analysis is the same as
the one described in the TPC analysis (subsection~\ref{TPC}), the
same tracking and feed-down corrections are applied. In the case
of the TOF analysis, an additional correction is needed in order
to take into account the fraction of the particles reconstructed
by the TPC with an associated signal in TOF. This matching
efficiency includes all sources of track losses in the propagation
from the TPC to the TOF (geometry, decays and interactions with
the material) and its matching with a TOF signal (the TOF
intrinsic detector efficiency, the effect of dead channels and the
efficiency of the track-TOF signal matching procedure). The TOF
matching efficiency has been derived from Monte-Carlo events as
the fraction of TPC reconstructed tracks having an associated TOF
signal and is shown in Fig.~\ref{TOF:matchingEff} for each hadron
species. The main factors limiting the TOF matching efficiency are
the loss due to geometrical acceptance ($\approx$ 15\%), the
number of dead or noisy channels ($\approx$ 10\%) and the
absorption of particles in the material of the transition
radiation detector ($\approx$ 8\%).

The TOF matching efficiency has been tested with data, using
\dedx\ in the TPC to identify the particles. Good agreement
between the efficiencies obtained from the data and from
Monte-Carlo simulations is observed in case of pions and kaons,
with deviations at the level of, at most, 3\% and 6\%
respectively, over the full transverse-momentum range.  The
observed differences are assigned as systematic errors, see Table
\ref{tab:tofsyst}. In the case of protons and antiprotons, larger
differences are observed at \pt\ below 0.7 GeV/$c$, where the
efficiency varies very rapidly with momentum. This region is
therefore not considered in the final results (see
Table~\ref{tab:tofsyst}).

Other sources of systematic errors related to the TOF PID
procedure have been estimated from Monte-Carlo simulations and
cross-checked with data. They include the effect of the residual
contribution from tracks wrongly associated with TOF signals, and
the quality and stability of the fit procedure used for extracting
the yields. Table~\ref{tab:tofsyst} summarizes the maximal value
of the systematic errors observed over the full transverse
momentum range relevant in the analysis, for each of the sources
mentioned above.

\begin{table}
\caption{Summary of systematic errors in the TOF analysis.}
\label{tab:tofsyst}
\begin{tabular}{cccc}
\hline\noalign{\smallskip}
systematic errors & $\pi^{\pm}$ & K$^{\pm}$ & p and \pbar \\
\noalign{\smallskip}\hline\noalign{\smallskip}

 TOF & $<$ 3\% & $<$ 6\% &  $<$ 4\% \\
    matching     &        &          & (\pt $> 1~{\rm GeV}/c$)\\
 efficiency             &          &          &  $<$ 7.5\%\\
   &        &          & (\pt $ = 0.7~{\rm GeV}/c$)\\
\noalign{\smallskip}\hline\noalign{\smallskip}

PID procedure & $<$ 2\%  & $<$ 7\%  & $<$ 3\%  \\
\noalign{\smallskip}\hline
\end{tabular}
\end{table}

\subsection{Kaon Identification using their decay within the TPC}
\label{sec:kinks}

In this section, the determination of the yields of charged kaons
identified by their weak decay (kink topology) inside the TPC
detector is described. These tracks are rejected in the previously
described TPC analysis. This procedure allows an extension of the
study of kaons to intermediate momenta, on a track-by-track level,
although in this analysis the \pt\ reach is limited by statistics.

The kinematics of the kink topology, measured as a secondary
vertex with one mother and one daughter track of the same charge,
allows the separation of kaon decays from the main source of
background kinks coming from charged pion decays. The decay
channels with the highest branching ratio (B.R.) for kaons are
the two-body decays
\begin{itemize}
\item [(1)] {$\rm{K^\pm}\rightarrow\mu^\pm +\nu_{\mu}$} ,  (B.R.
63.55\%) \item [(2)] {$\rm{K^\pm}\rightarrow\pi^\pm + \pi^0$}  ,
(B.R. 20.66\%).
\end{itemize}

Three-body decays with one charged daughter track (B.R. 9.87\%) as
well as three-body decays into three char\-ged pions (B.R. 5.6\%)
are also detected.

The  algorithm for reconstructing kinks as secondary vertices is
applied inside a fiducial volume of the TPC with radius {120 cm $<
R<$ 210 cm} in order to have a minimum number of clusters for
reconstructing both the mother and daughter tracks. Inside this
volume  a sufficient number of kinks can be found since the
$c\tau$ of kaon and
 pion decays  are  3.7 m and 7.8 m, respectively.
The mother track of the kink has been selected with similar
criteria to those of the TPC tracks used for the \dedx\ analysis,
except that the minimum required number of clusters per  track is
30, because the kink mother track does not traverse the entire
TPC. The relation between the number of clusters per mother track
and the radius $R$ of the kink is used as a quality check of the
kink reconstruction procedure. 

\begin{figure}[h]
\begin{center}
\resizebox{0.36\textwidth}{!}{%
\includegraphics{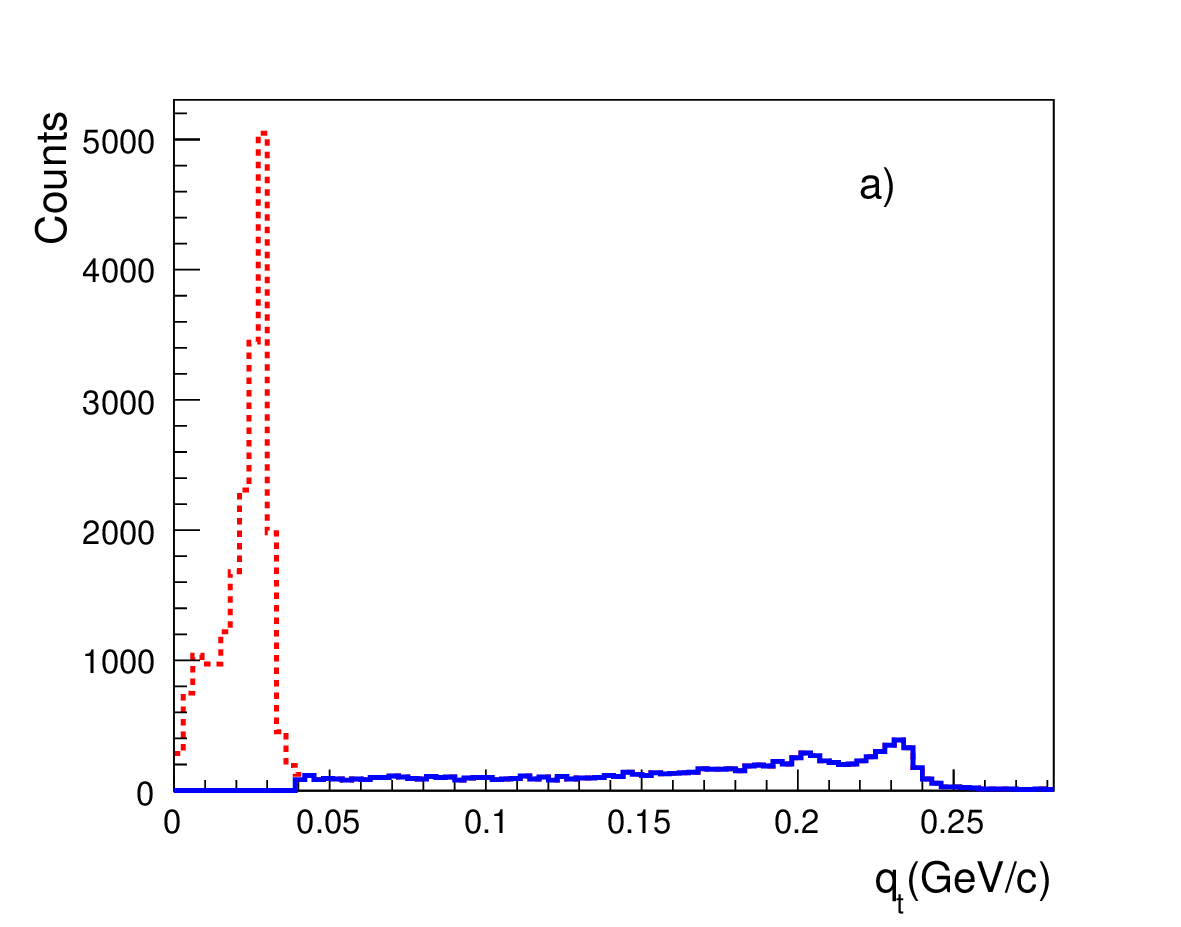}
}
\resizebox{0.36\textwidth}{!}{%
\includegraphics{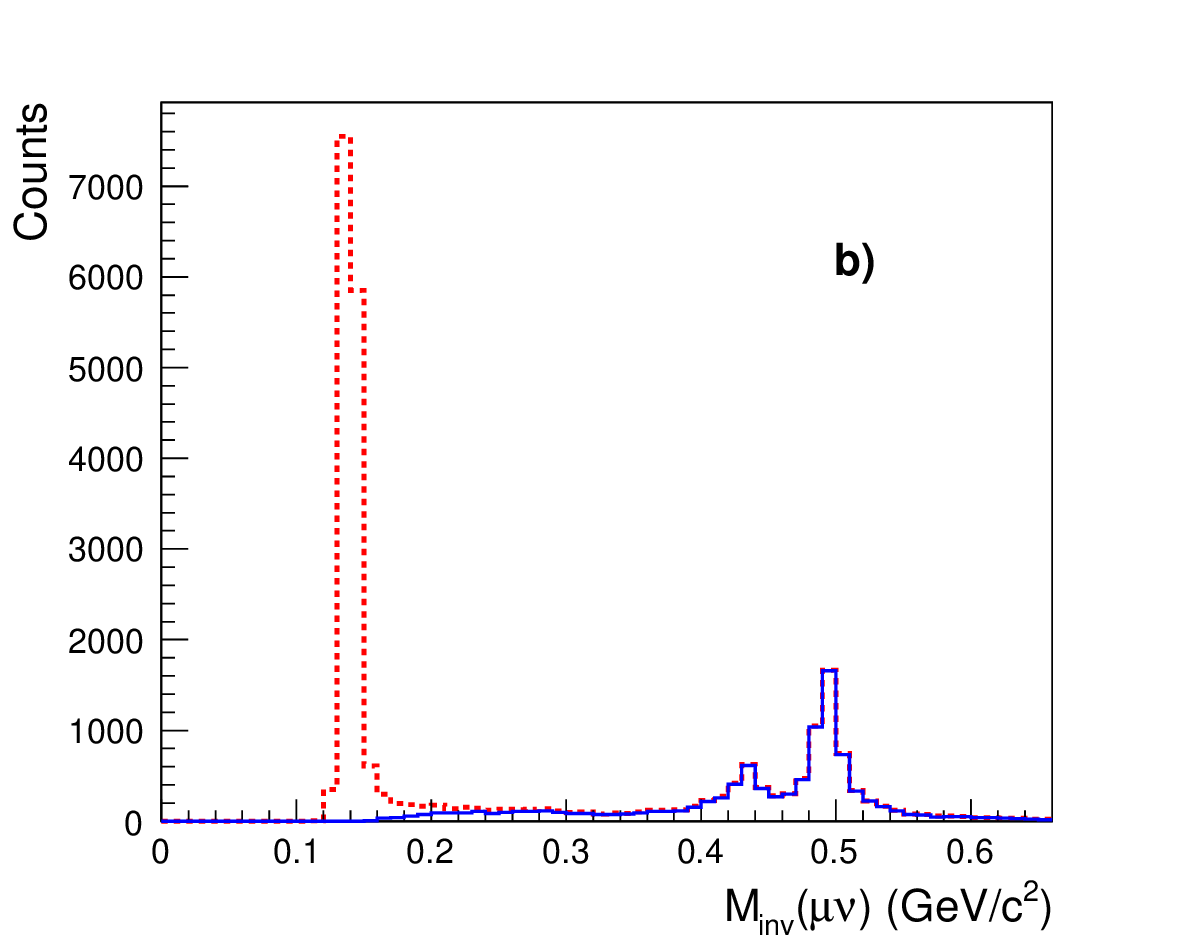}
}
\resizebox{0.36\textwidth}{!}{%
\includegraphics{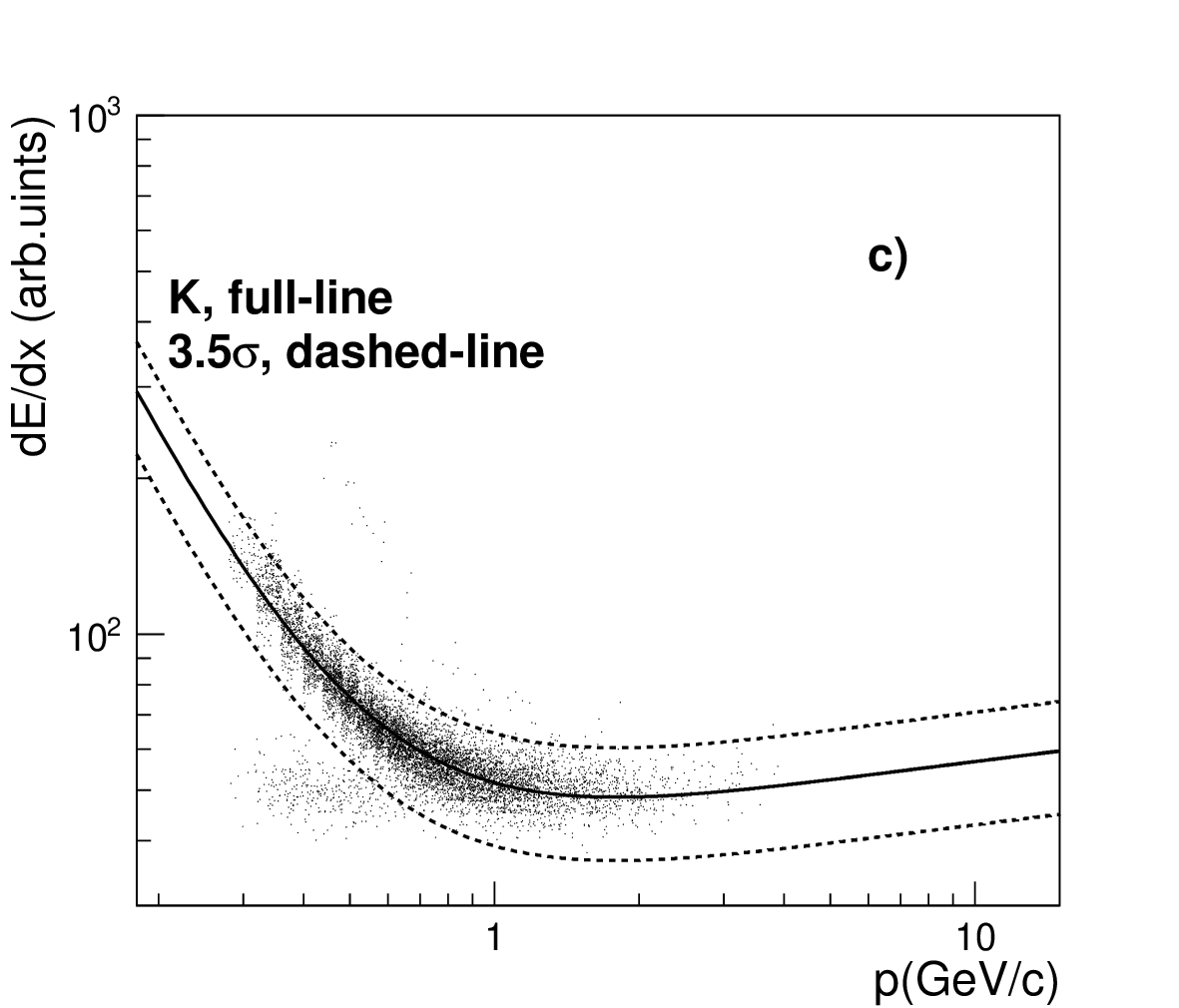}
} \caption{(Color online) (a) $q_{\rm t}$ distribution of the
daughter tracks with respect to mother momentum for all
reconstructed kinks inside the analyzed sample. The dashed(solid)
histograms show the distribution before (after) applying the
$q_{\rm t}
> 40$~MeV/$c$ cut.
(b) Invariant mass of the two-body decays
$\rm{K^\pm}/{\pi^\pm}\rightarrow\mu^\pm +\nu_{\mu}$ for candidate
kaon kinks. Solid curve: after applying $q_{\rm t}>$40 MeV/$c$;
dashed curve: without this selection (hence also showing the pion
decays).
(c) \dedx\ of kinks as a function of the mother momentum, after
applying the full list of selection criteria for their
identification.
 } \label{fig:KinkSelection}
\end{center}
\end{figure}

The identification of  kaons from kink topology and its separation
from pion decay is based on the decay kinematics. The transverse
momentum of the daughter with respect to the mother's direction,
$q_{\rm t}$, has  an upper limit of 236 MeV/$c$ for kaons and 30
MeV/$c$ for  pions for the two-body decay to $\mu +\nu_{\mu}$. The
corresponding upper limit for the two-body decay (2)
{$\rm{K}\rightarrow\pi +\pi^{0}$} is 205 MeV/$c$. All three limits
can be seen as peaks in Fig.~\ref{fig:KinkSelection} (a), which
shows the $q_{\rm t}$ distribution of all measured kinks inside
the selected volume and rapidity range $|y|$ $<$ 0.7. Selecting
kinks with $q_{\rm t} > 40$ MeV/$c$ removes the majority of
$\pi$-decays as shown by the dashed (before) and solid (after)
histograms.

The invariant mass for the decay into $\mu^\pm +\nu_{\mu}$ is
calculated from the measured difference between the mother and
daughter momentum, their decay angle, assuming zero mass for the
neutrino. Figure~\ref{fig:KinkSelection} (b) shows the invariant
mass for the full sample of kinks (dashed line) and for the sample
after applying the preceding cuts (full line). 
The masses of pions and   kaons are reconstructed at their nominal
values. The third  peak at 0.43 GeV/$c$ originates from the
{$\rm{K}\rightarrow\pi + \pi^0$} decay for which the invariant
mass is calculated with wrong mass assumptions for the daughter
tracks. The broad structure originates from three-body decays of
kaons.

At this stage, we have a rather clean sample of kaons as
demonstrated in Fig.~\ref{fig:KinkSelection} (c) showing the
\dedx\ vs.~the mother momentum. Most of the tracks  are within a
$3.5\sigma$ band  with respect to the corresponding  Bethe-Bloch
curve of kaons. The  few tracks outside  these limits are at
momenta below 600 MeV/$c$ (less than 5\%) and they have been
removed in the last analysis step.

\paragraph{Efficiency and acceptance}

The total correction factor includes both the acceptance of kinks
and their efficiency (reconstruction and identification). The
study has been performed for the rapidity interval $|y|$ $<$ 0.7,
larger than the corresponding rapidity interval for the other
studies
  in order to   reduce the statistical errors.

\begin{figure}
\begin{center}
\resizebox{0.45\textwidth}{!}{%
\includegraphics{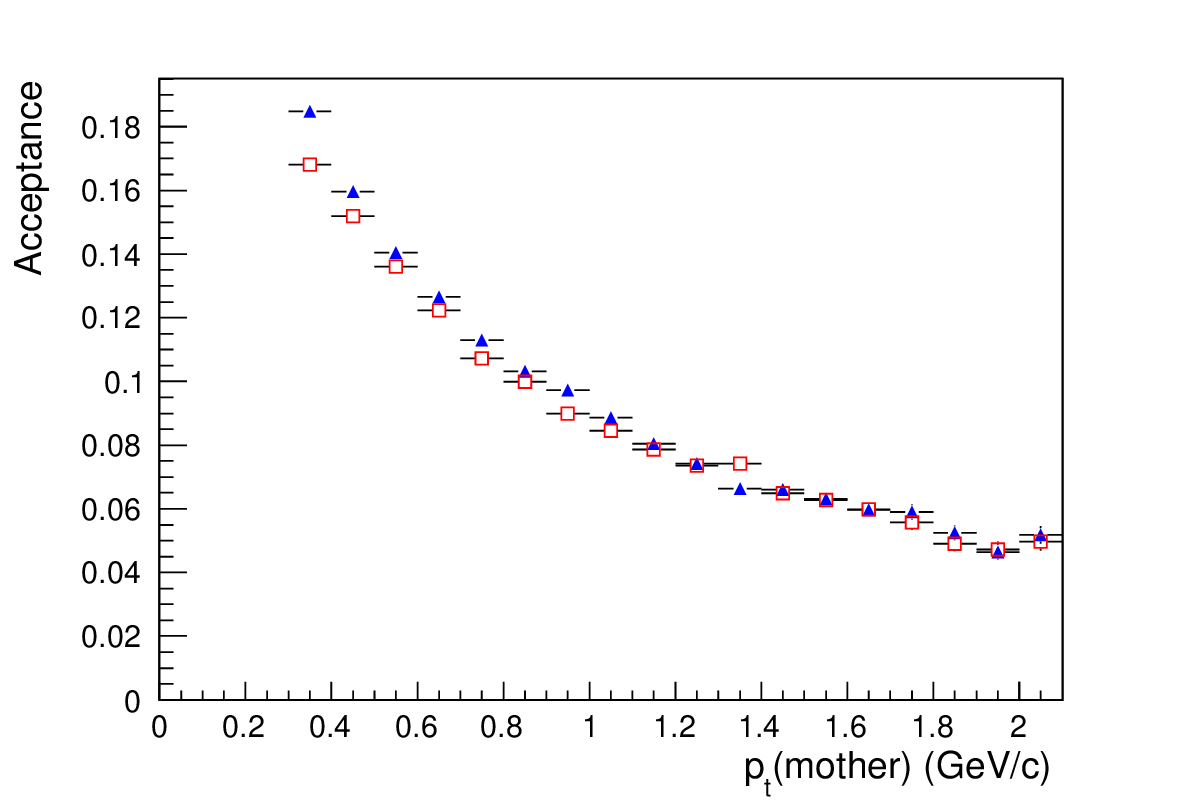}
}
\resizebox{0.45\textwidth}{!}{%
\includegraphics{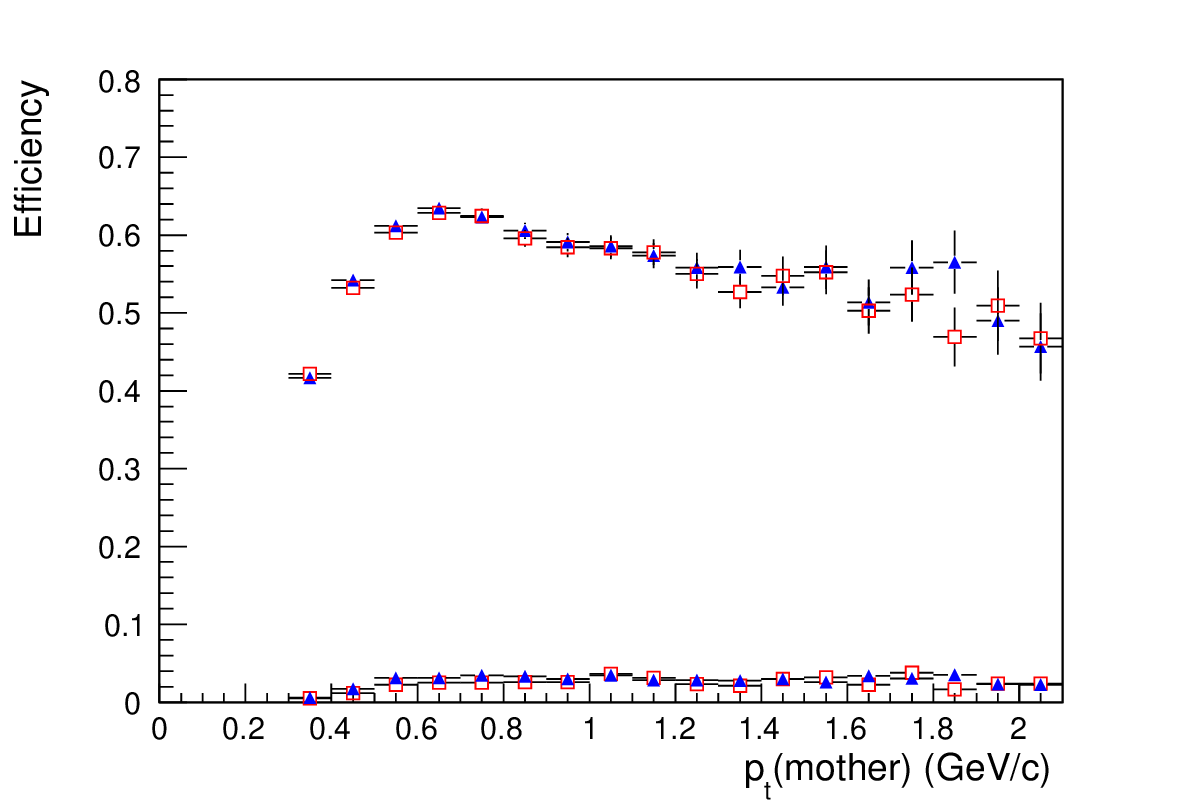}
} \caption{(Color online) Upper panel: The acceptance of kaons
decaying in the fiducial volume of the TPC as a function of the
kaon \pt for $\rm{K}^+$ (full-triangles)  and $\rm{K}^-$
(open-squares). Lower panel: The  efficiency of reconstructed
kaons from kinks as a function of the \pt\ (mother), separately
for $\rm{K}^+$ (full-triangles) and $\rm{K}^-$ (open-squares). The
contamination from wrongly associated kinks is also plotted for
both charges (lower set of points). }\label{fig:kink_eff}
\end{center}
\end{figure}

The acceptance is  defined as the ratio of weak decays (two- and
three-body decays) whose daughters are inside the fiducial volume
of the TPC to all kaons inside the same rapidity window
(Fig.~\ref{fig:kink_eff}, upper part). It essentially reflects the
decay probability. However, the acceptance is not the same in the
low-momentum region
 for both charges of kaons, since the  interaction cross
section of the negative kaons with the  ITS material  is higher
than that of the positive kaons. As a result, the acceptance of
positive kaons is larger at low momenta.

The efficiency is the ratio of reconstructed and identified kaons
divided by the number of kaon decays within the acceptance as
shown in Fig.~\ref{fig:kink_eff} (lower part), as a function of
the kaon \pt. It reaches about 60\%  at 0.7 GeV/$c$ and decreases
gradually at higher transverse momenta, as the angle between
mother and daughter tracks becomes  smaller. The decay angle of
kaon kinks allows their identification up to high momenta, e.g.~at
\pt\ of 5 GeV/$c$ the values are between $2^\circ$ and $15^\circ$.

The contamination due to random associations  of primary and
secondary charged tracks 
 has been established using Monte-Carlo
simulations and it is systematically smaller than 5\% in the
studied  \pt-range as also shown in Fig.~\ref{fig:kink_eff}.
Hadronic interactions are the main source of these fake kinks
(65\%).

The systematic error due to the uncertainty in the material budget
is about 1\% as for the TPC analysis. The quality cuts remove
about 8\%  of all real kaon kinks, which leads to a systematic
error of less than 1\%. The main uncertainty originates from the
efficiency of the kink finding algorithm which has an uncertainty
of 5\%.

\section{Results}
\label{results}

\begin{figure}[tbp]
\resizebox{0.48\textwidth}{!}{%
\includegraphics{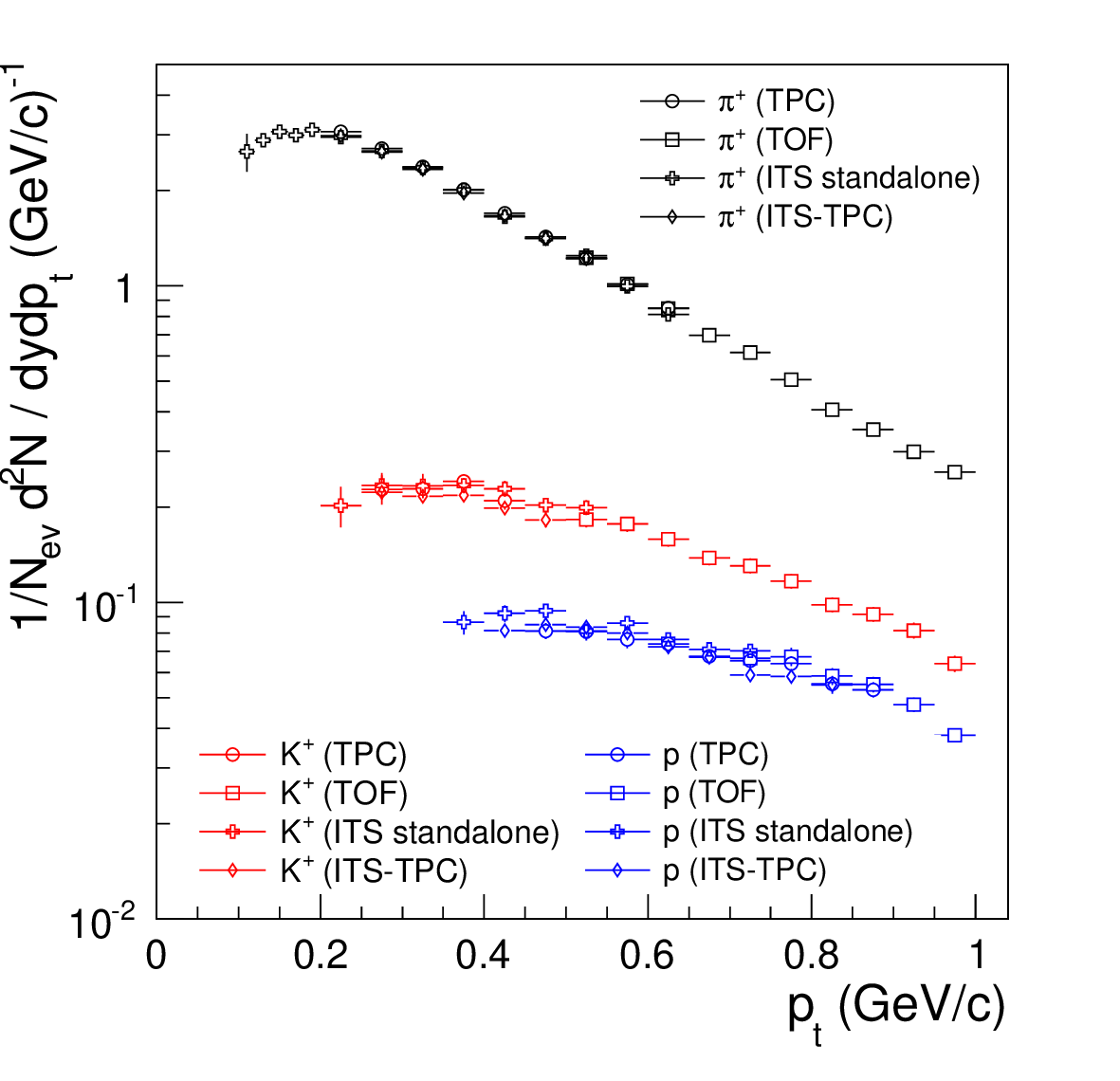}
}
\resizebox{0.48\textwidth}{!}{%
\includegraphics{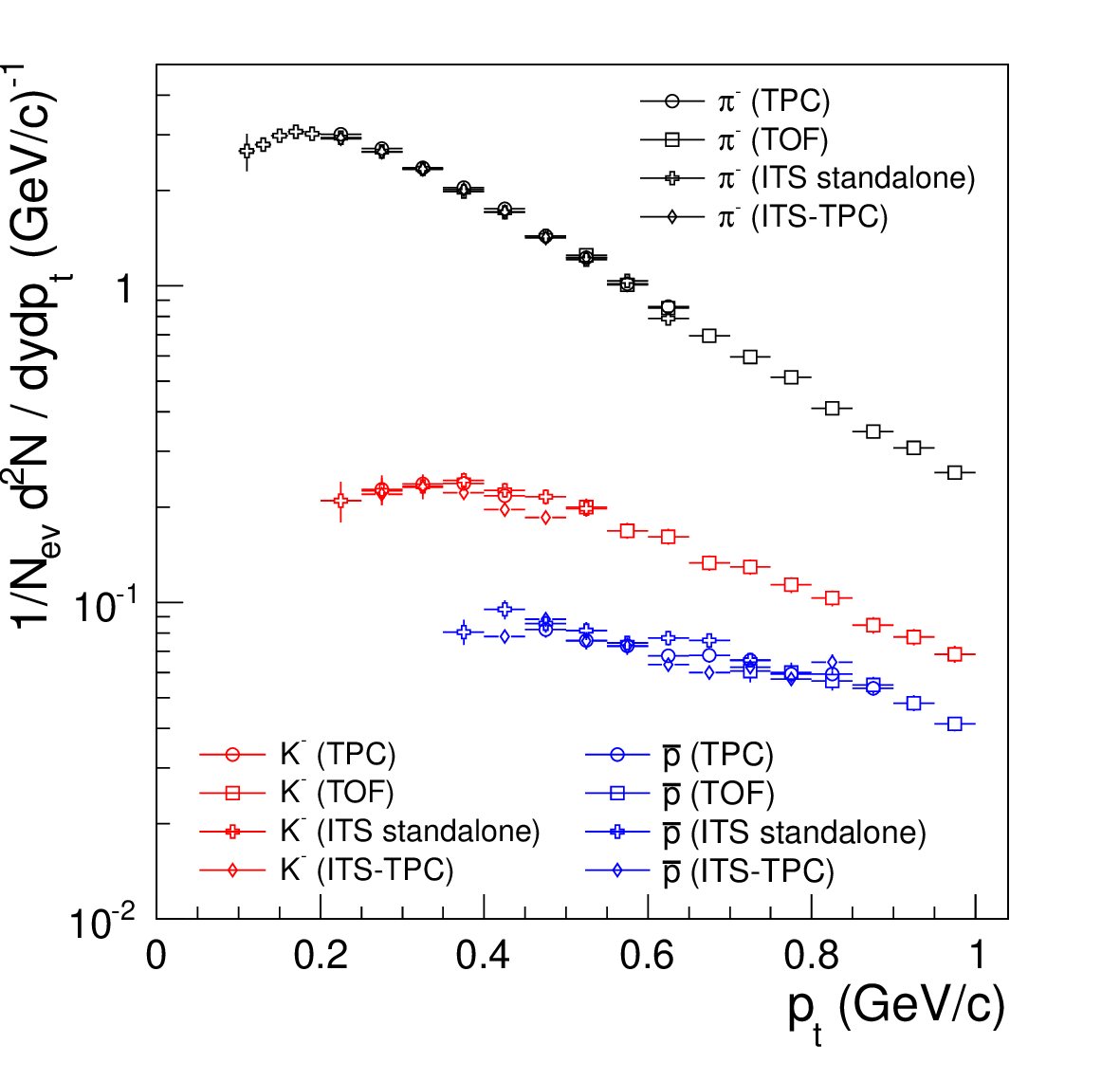}
}
 \caption{(Color online) Transverse momentum spectra ${\rm d}^2N / ({\rm d}p_{\rm t}{\rm d}y)$
 for $|y|<$ 0.5 of positive (upper part) and negative (lower part)
 hadrons from the various analyses. Only systematic errors are plotted.}
 \label{fig:spectra_det}
\end{figure}

\begin{figure}[tbp]
\resizebox{0.45\textwidth}{!}{%
\includegraphics{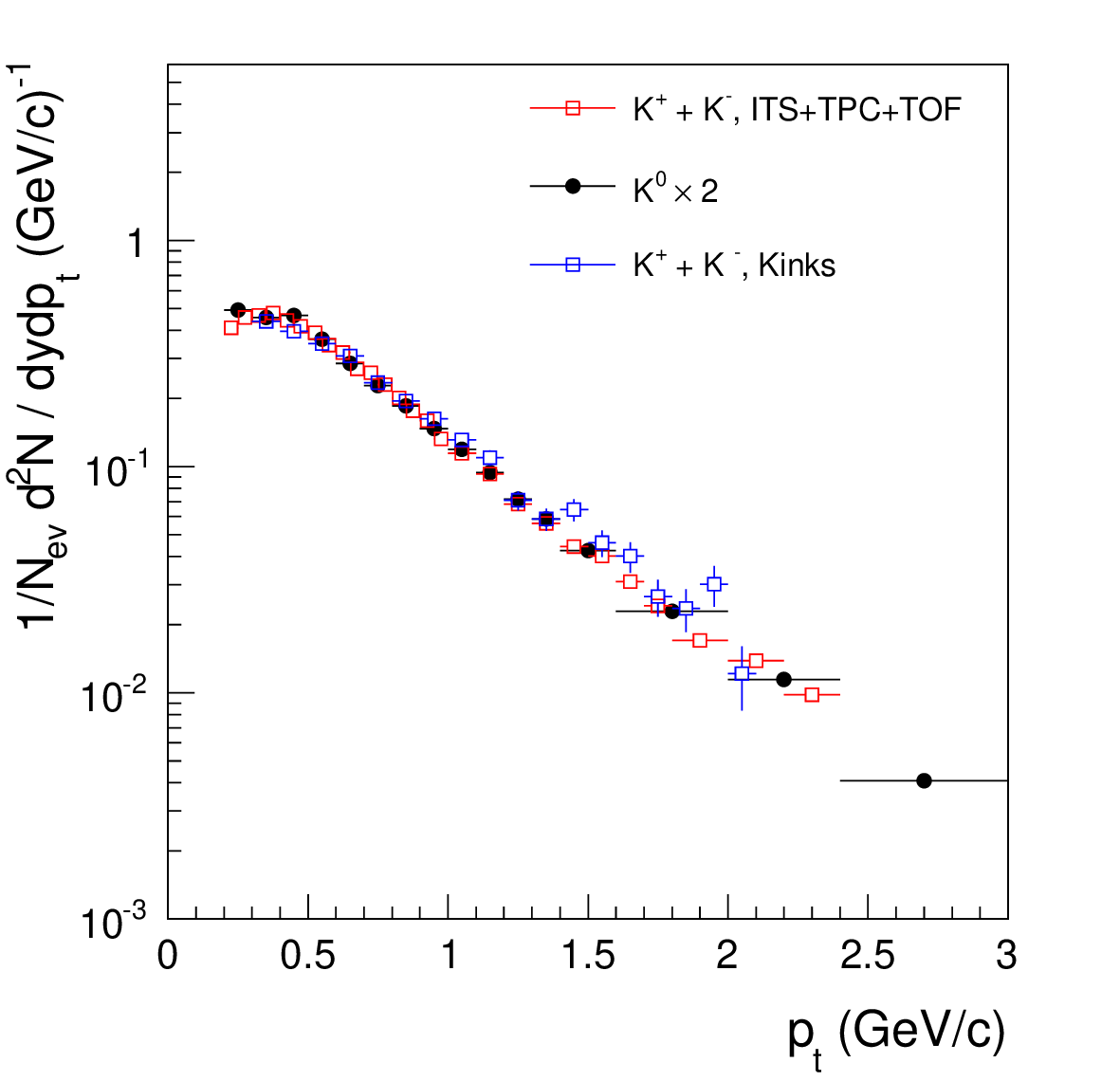}
} \caption{(Color online) Comparison of charged kaon spectra,
obtained from the combined ITS stand-alone, TPC, TOF analysis,
from the kink topology and K$^0_s$ spectra from
Ref.~\cite{strange}}. Only statistical errors are shown.
\label{fig:spectra_kaons}
\end{figure}

Figure~\ref{fig:spectra_det} shows a comparison between the
results from the different analyses.  The spectra are normalized
to inelastic collisions, as explained in Sec.~\ref{events}.  The
kaon spectra obtained with various techniques, including 
K$^0_s$ spectra~\cite{strange}, are compared in
Fig.~\ref{fig:spectra_kaons}. The very good agreement demonstrates
that all the relevant efficiencies are well reproduced by the detector simulation.

The spectra from ITS stand-alone, TPC and TOF are combined in
order to cover the full momentum range. The analyses from the
different detectors use a slightly different sample of tracks and
have largely independent systematics (mainly coming from the PID
method and the contamination from secondaries). The spectra have
been averaged, using the systematic errors as weights.  From this
weighted average, the combined, \pt-dependent, systematic error is
derived. The combined spectra have an additional overall
normalization error, coming primarily from the uncertainty on the
material budget (3\%, Sec.~\ref{sec:part-ident}) and from the
normalization procedure (2\%, Sec.~\ref{events}).

\begin{figure}[h]
\resizebox{0.45\textwidth}{!}{%
\includegraphics{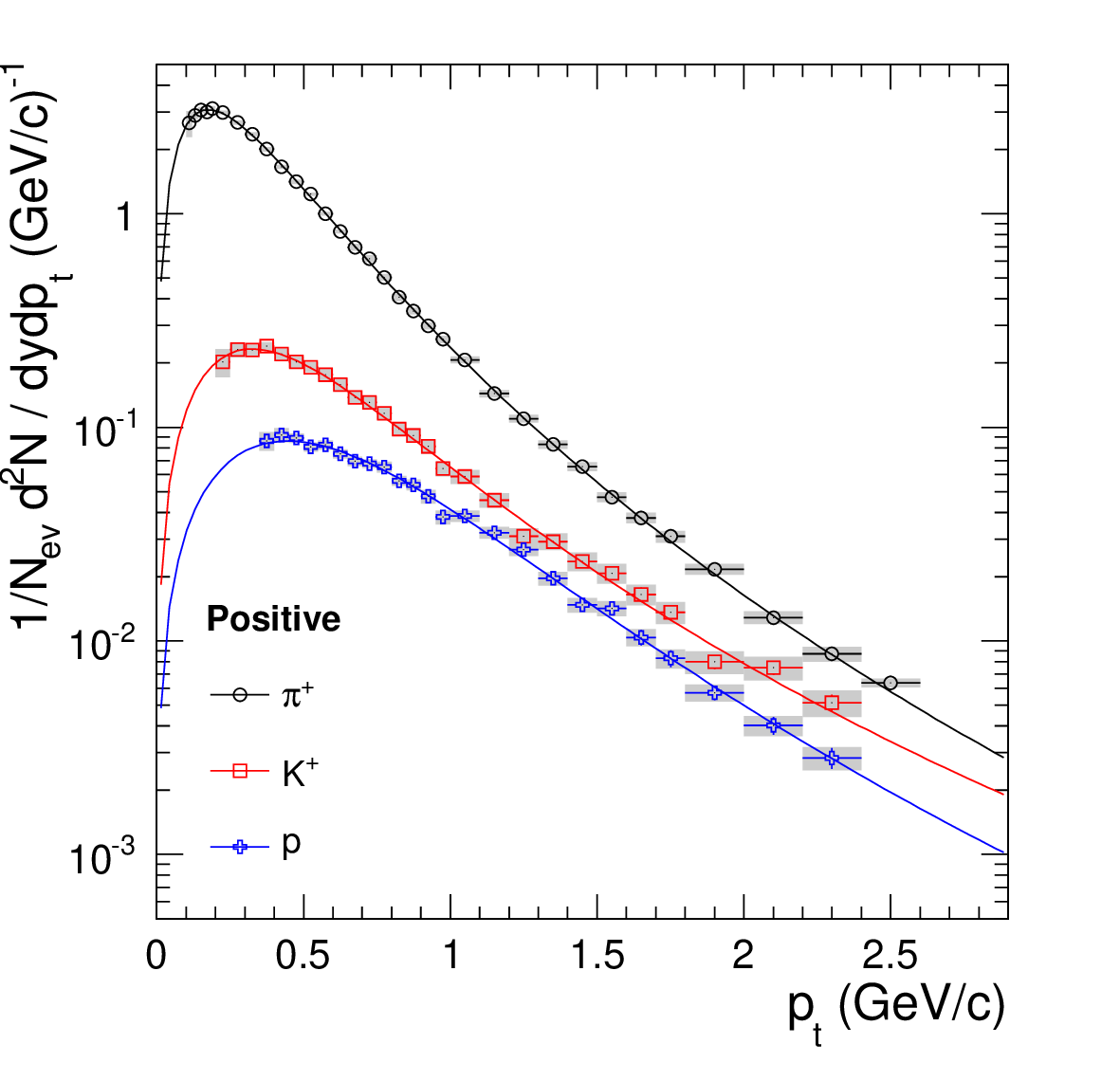}
}
\resizebox{0.45\textwidth}{!}{%
\includegraphics{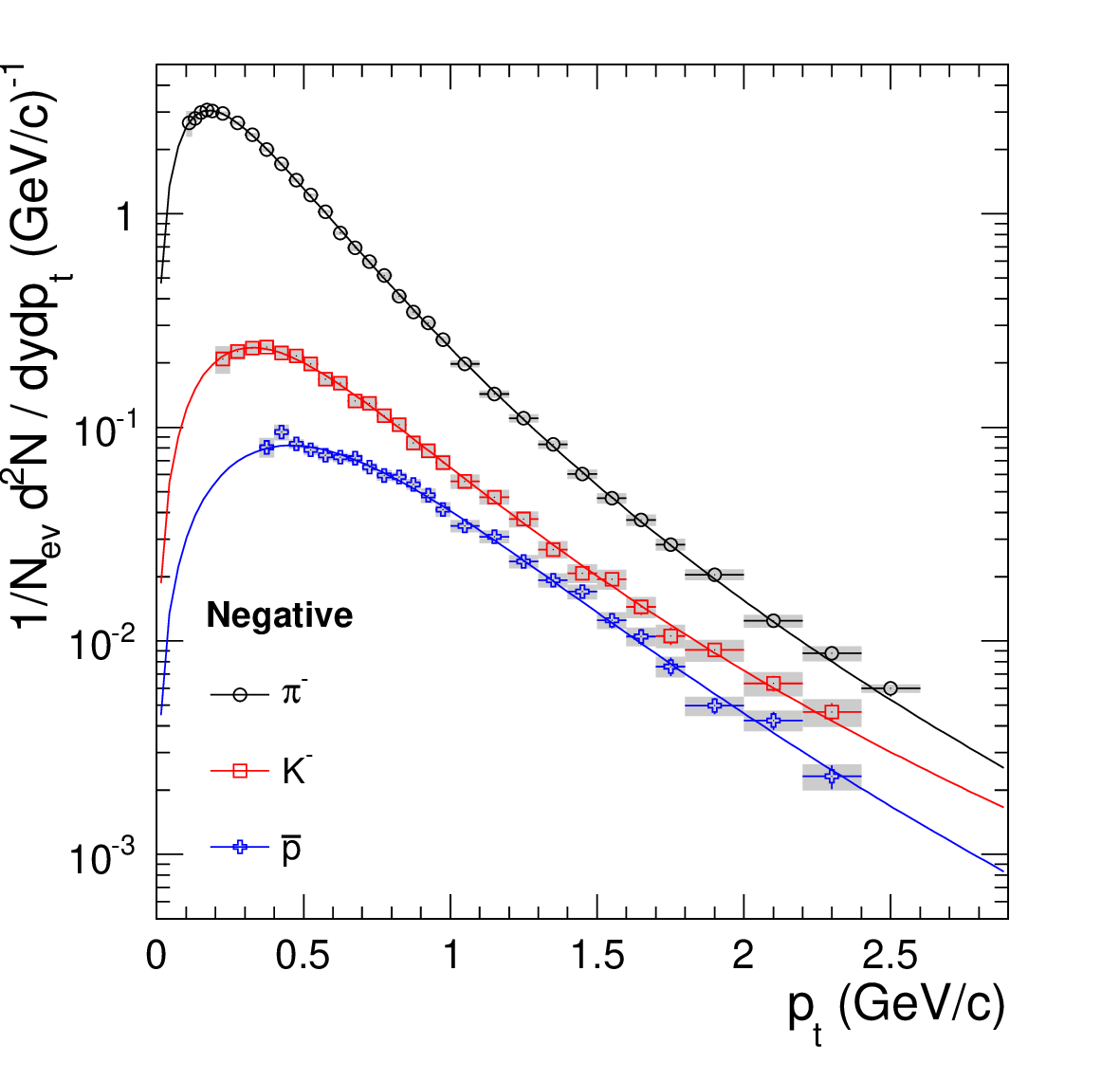}
} \caption{(Color online) Transverse momentum spectra of positive
(top) and negative (bottom) hadrons from pp collisions at \s = 900
GeV. Grey bands: total \pt-dependent error (systematic plus
statistical); normalization systematic error (3.6\%) not shown.
The curves represent fits using a L\'{e}vy function. }
\label{fig:spectra}
\end{figure}

The combined spectra shown in Fig.~\ref{fig:spectra} are fitted
with the L\'{e}vy (or Tsallis) function (see
e.g.~\cite{Tsallis:1987eu,Abelev:2006cs})
\begin{equation}
  \label{eq:1}
  \frac{{\rm d}^2N}{{\rm d}p_{\rm t}{\rm d}y} = p_{\rm t} \times \frac{{\rm d}N}
  {{\rm d}y} \frac{(n-1)(n-2)}{nC(nC + m_{0} (n-2))} \left( 1 + \frac{m_{\rm t} - m_{0}}{nC} \right)^{-n}
\end{equation}

with the fit parameters $C$, $n$ and the yield ${\rm d}N/{\rm
  d}y$. This function gives a good description of the spectra and has
been used to extract the total yields and the $\langle p_{\rm t}
\rangle$, summarized in Table~\ref{tab:levy}. The
$\chi^{2}$/degree-of-freedom is calculated using the total error.
Due to residual correlations in the point-by-point systematic
error, the values are less than 1.
Also listed are 
the lowest measured \pt-bin and the fraction of the yield
contained in the extrapolation of the spectra to zero momentum.
The extrapolation to infinite momentum gives a negligible
contribution. The systematic errors take into account the
contributions from the individual detectors, propagated to the
combined spectra, the overall normalization error and the
uncertainty in the extrapolation. The latter is evaluated using
different fit functions (modified Hagedorn~\cite{Hagedorn:1983wk}
and the UA1 parametrization~\cite{Albajar:1989an}) or using a
Monte-Carlo generator, matched to the data for $p_{\rm t} <
1~\mathrm{GeV}/c$ (PYTHIA~\cite{Sjostrand:2006za}, with tunes
D6T~\cite{D6T}, CSC and Perugia0~\cite{perugia}, or
PHOJET~\cite{Engel:1995sb}). While none of these alternative
extrapolations provides a description as good as the one from the
L\'{e}vy fit, we estimate from this procedure an uncertainty of
about 25\% of the extrapolated part of the yield.

The ratios of \pip/\pim and \kap/\kam\ as a function of \pt\ are
close to unity within the errors, allowing the combination of both
spectra in the L\'{e}vy fits.
The \pbar/p ratio as a function of \pt\ has been studied with high
precision in our previous publication~\cite{Aamodt:2010dx}. It is
\pt-independent with a mean value of
$0.957\pm0.006(\rm{stat})\pm0.014 (\rm{syst})$. Also here we used
the sum of both charges. Table~\ref{tab:levy_values} summarizes
the fit parameters along with the yields and mean \pt. The errors
have been determined as for the individual fits.

\begin{table*}
\caption{Integrated yield d$N$/d$y$ ($|y|<0.5$) with statistical
and systematic errors, and $\langle p_{\rm t} \rangle$, as
obtained from the fit with the L\'{e}vy function together with the
lowest \pt\ experimentally accessible, the fraction of
extrapolated yield and the $\chi^2$/ndf of the fit (see text). The
systematic error of d$N$/d$y$ and of the $\langle p_{\rm t}
\rangle$ includes the contributions from the systematic errors of
the individual detectors, from the choice of the functional form
for extrapolation and from the absolute
normalization. 
} \center
\label{tab:levy}       
\begin{tabular}{cccccccc}
\hline\noalign{\smallskip}
 Particle   & d$N$/d$y$                    & $\langle$ \pt\ $\rangle$  (GeV/$c$)      & Lowest \pt\ (GeV/$c$)  & Extrapolation & $\chi^2$/ndf \\
\hline\noalign{\smallskip}
$\pi^{+}$   &  $1.493\pm 0.004\pm 0.074$ &  $0.404\pm 0.001\pm$ 0.02   & $0.10$  & 10\% &   14.23/30 \\
$\pi^{-}$   &  $1.485\pm 0.004\pm 0.074$ &  $0.404\pm 0.001\pm$ 0.02   & $0.10$  & 10\% &   12.46/30 \\
K$^{+}$     &  $0.183\pm 0.004\pm 0.015 $ &  $0.658\pm 0.006\pm$ 0.05  & $0.20$  & 13\% &   12.71/24 \\
K$^{-}$     &  $0.182\pm 0.004\pm 0.015 $ &  $0.642\pm 0.006\pm$ 0.05  & $0.20$  & 13\% &   6.23/24 \\
p           &  $0.083\pm 0.002\pm 0.006$ &  $0.768\pm 0.008\pm$ 0.06    & $0.35$  & 21\% &   13.79/21 \\
\pbar       &  $0.079\pm 0.002\pm 0.006$ &  $0.760\pm 0.008\pm$ 0.06    & $0.35$  & 21\% &   13.46/21 \\
\end{tabular}
\end{table*}

\begin{table*}
\caption{Results of the L\'{e}vy fits to combined positive and
negative spectra. See text and the caption of Table~\ref{tab:levy}
for details on the systematic errors.} \center
\begin{tabular}{cccccccc}
\hline\noalign{\smallskip}
Particle & d$N$/d$y$ &  $C$ (GeV) & $n$  & $\langle$ \pt\ $\rangle$ (GeV/$c$)& $\chi^2$/ndf \\
\hline\noalign{\smallskip}
$\pi^{+}$ + $\pi^{-}$ & $2.977  \pm 0.007\pm   0.15$  &$0.126 \pm 0.0005 \pm 0.001$&  $7.82 \pm 0.06\pm 0.1$   &$0.404 \pm 0.001 \pm 0.02$&19.69/30 \\
K$^{+}$ + K$^{-}$    & $0.366  \pm 0.006\pm   0.03$  &$0.160  \pm 0.003  \pm 0.005$&   $6.08 \pm 0.2  \pm 0.4$  &$0.651 \pm 0.004 \pm 0.05$&8.46/24  \\
p + \pbar           &  $0.162 \pm 0.003\pm 0.012$  &$0.184  \pm 0.005  \pm 0.007$&    $7.5 \pm 0.7  \pm 0.9$ & $0.764 \pm 0.005 \pm 0.07$&15.70/21 \\
\label{tab:levy_values}
\end{tabular}
\end{table*}

\begin{figure}[th!]
\resizebox{0.45\textwidth}{!}{%
\includegraphics{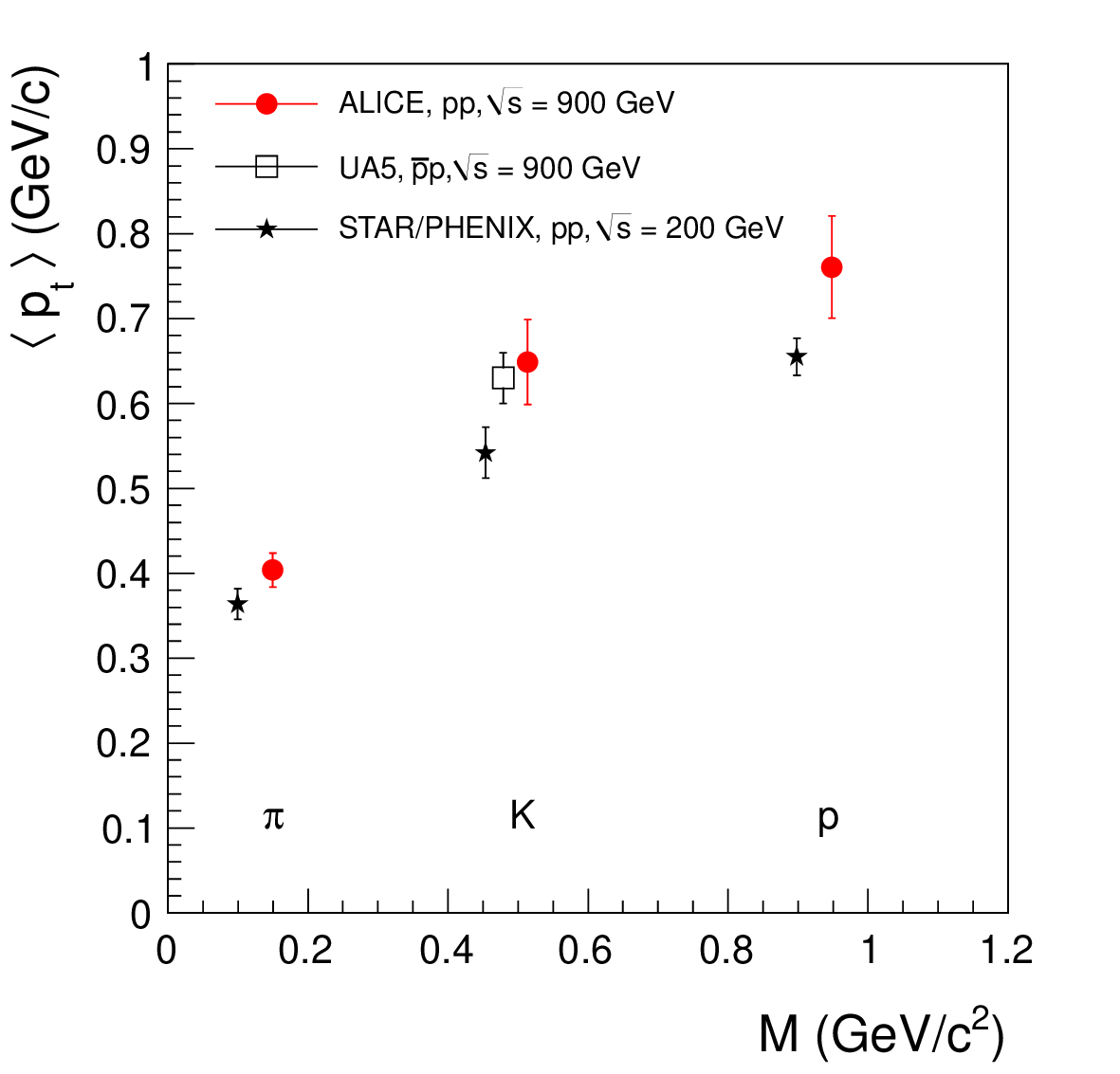}
} \caption{(Color online) Mean \pt\ as a function of the mass of
the emitted particle in pp collisions at 900 GeV (ALICE, red solid
circles, statistical and systematic errors) compared to results at
\s\ = 200 GeV (star markers, average values of the results from
the STAR and the PHENIX
Collaborations~\cite{:2008ez,Adare:2010fe}) and \pbar p reactions
at \s\ = 900 GeV~\cite{Ansorge:1987cj} (open squares). Some data
points are displaced for clarity.} \label{fig:mean_pt}
\end{figure}

\begin{figure}[th]
\resizebox{0.45\textwidth}{!}{%
\includegraphics{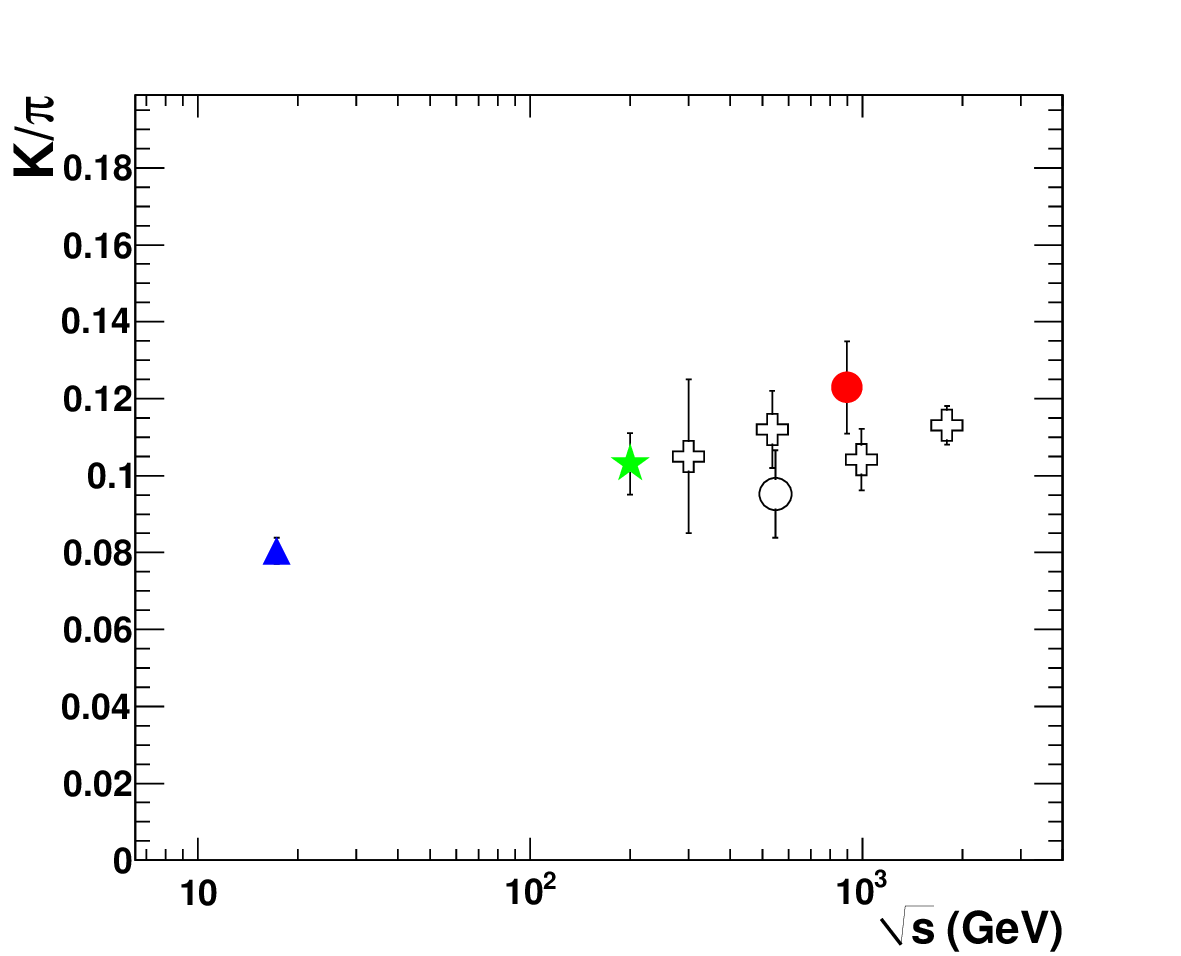}
} \caption{(Color online) Ratios (\kap+\kam)/(\pip+\pim) and
K$^0/\pi$ as a function of \s. Data (full symbols) are from pp
collisions, (at \s\ = 17.9 GeV by
NA49~\cite{Alt:2005zq,Anticic:2010yg}, at \s\ = 200 GeV by
STAR~\cite{:2008ez} and at \s\ = 900 ALICE, present work) and
(open symbols) from \pbar p interaction (at \s\ = 560 GeV by
UA5~\cite{Alner:1985ra} and at the TEVATRON by
E735~\cite{Alexopoulos:1993wt,Alexopoulos:1992ut}).}
\label{fig:Ktopi}
\end{figure}

\begin{figure}[th]
\resizebox{0.4\textwidth}{!}{%
\includegraphics{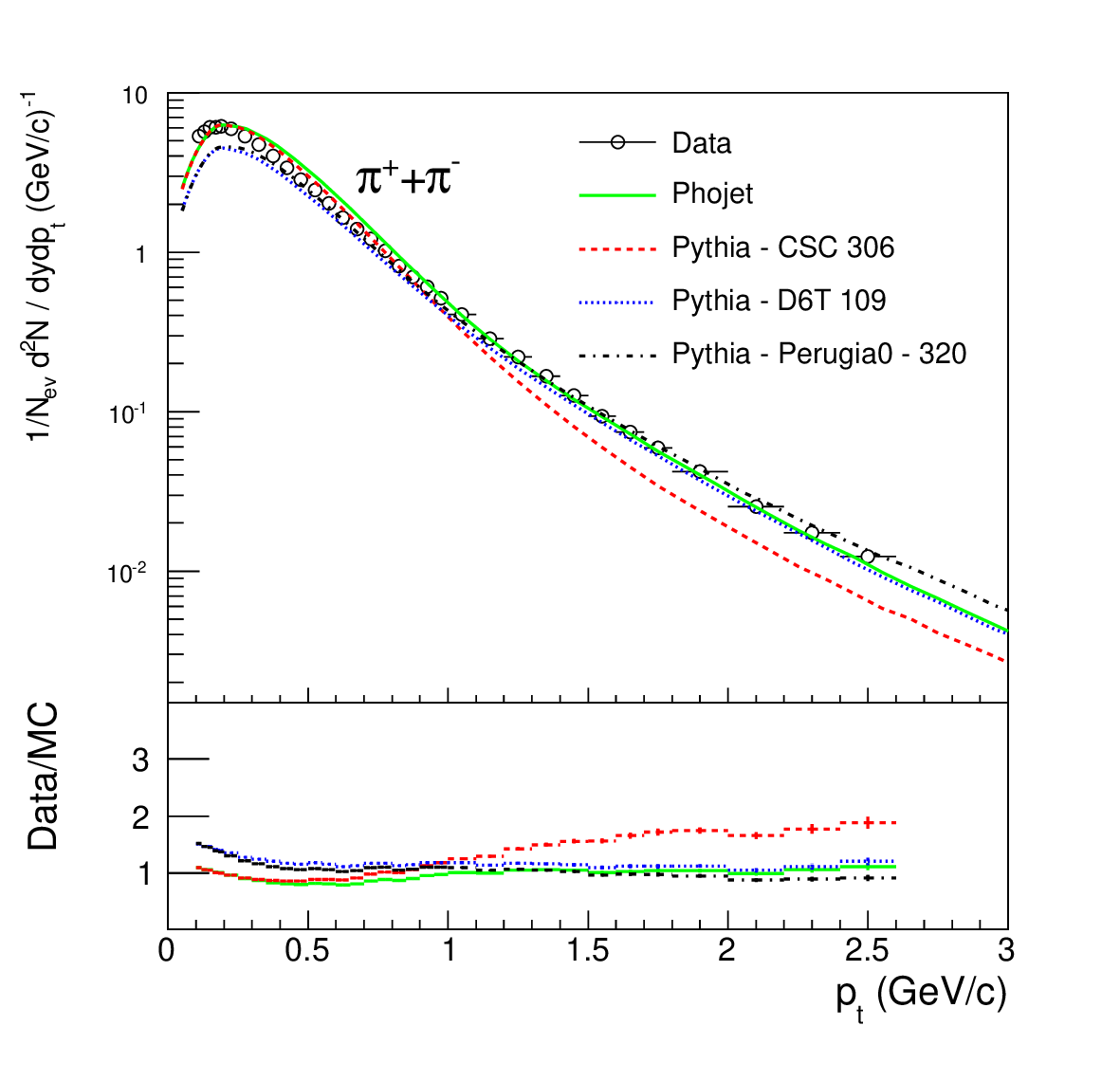}
}
\resizebox{0.4\textwidth}{!}{%
\includegraphics{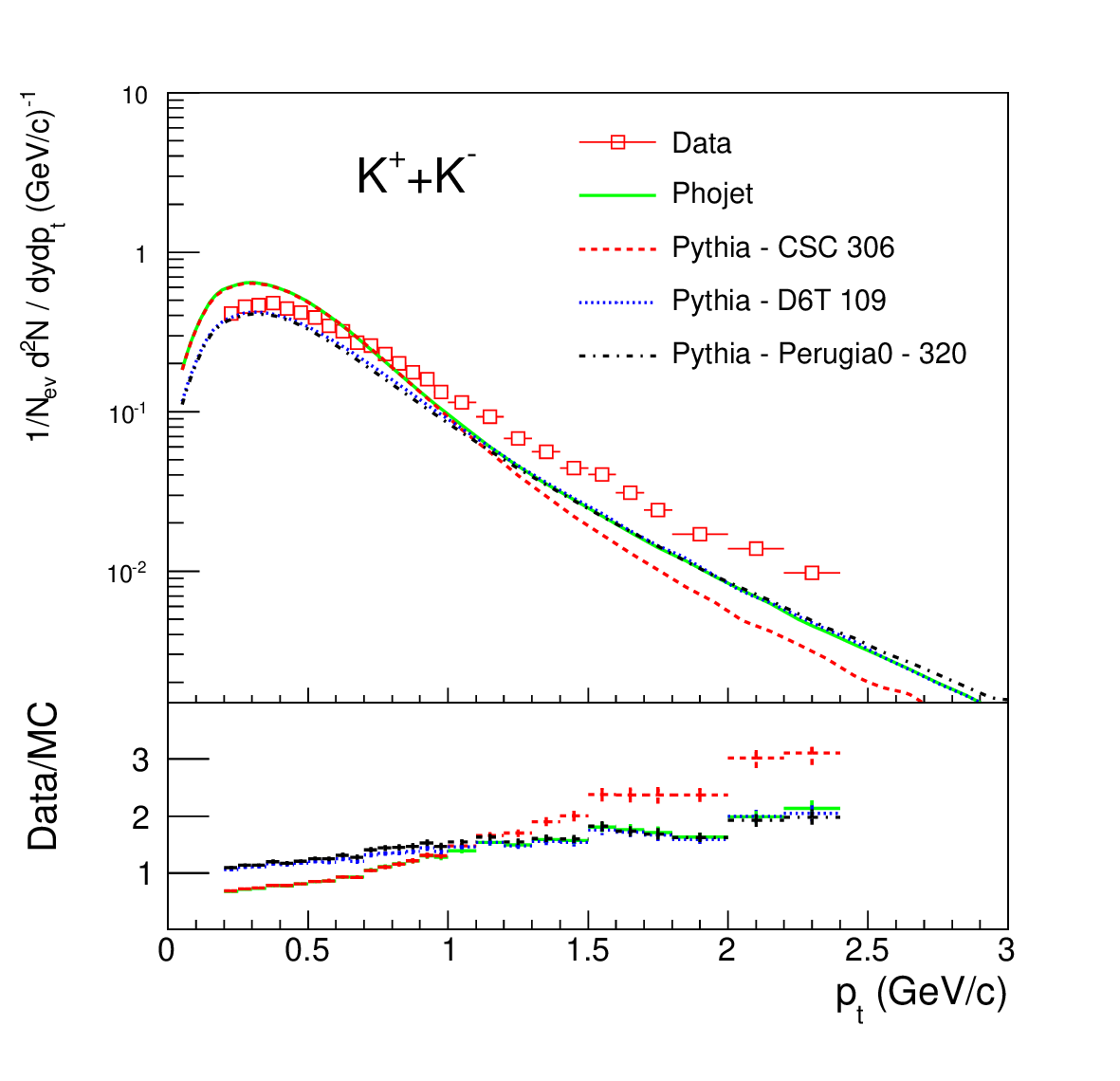}
}
\resizebox{0.4\textwidth}{!}{%
\includegraphics{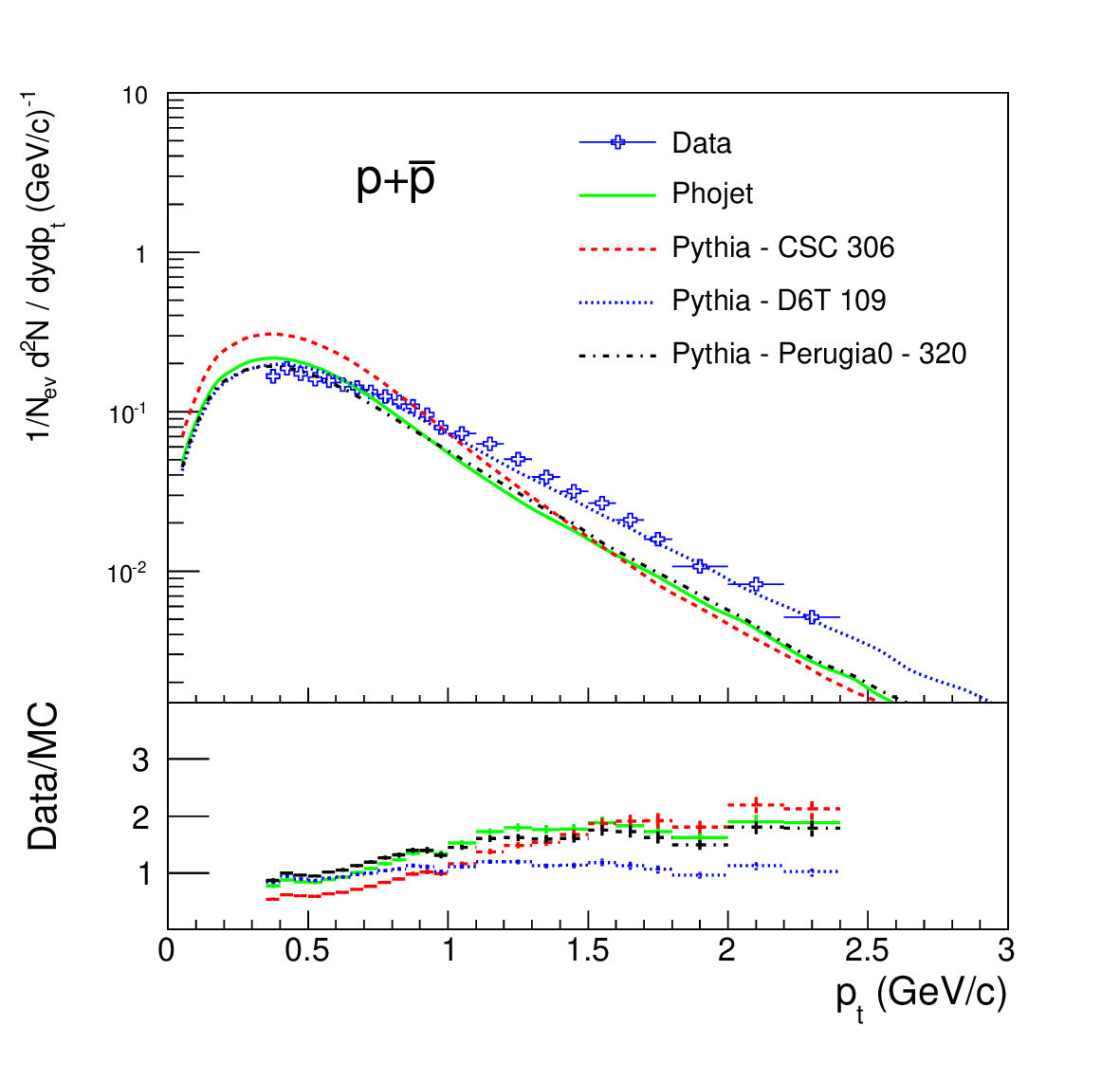}
} \caption{(Color online) Comparison of measured pion, kaon and
proton spectra at \s\ = 900 GeV (both charges combined) with
various tunes of event generators. Statistical errors only. See
text for details.} \label{fig:pythia-pos}
\end{figure}


Our values on yield and $\langle p_{\rm t} \rangle$ given in
Table~\ref{tab:levy} and \ref{tab:levy_values} agree well with the
results from \pbar p collisions at the same
\s~\cite{Ansorge:1987cj}. Figure~\ref{fig:mean_pt} compares the
$\langle p_{\rm t} \rangle$ with measurements in pp collisions at
\s\ = 200 GeV~\cite{:2008ez,Adare:2010fe} and in \pbar p reactions
at \s\ = 900 GeV~\cite{Ansorge:1987cj}. The mean \pt\ rises very
little with increasing \s\ despite the fact that the spectral
shape clearly shows  an increasing contribution from hard
processes. It was already observed at RHIC that the increase in
mean \pt\ at \s = 200 GeV compared to studies at \s = 25 GeV is
small.
The values obtained in pp collisions are lower than those for
central Au+Au reactions at \s = 200 GeV~\cite{:2008ez}.

The spectra presented in this paper are normalized to inelastic
events. In a similar study by the STAR Collaboration the yields
have been normalized to NSD collisions~\cite{:2008ez}. In order to
compare these two results, the yields in Table~\ref{tab:levy} have
been scaled to NSD events, multiplying by 1.185 (see Section 2.2).
The yields of pions increase from \s = 200 GeV to 900 GeV by 23\%,
while \kap\ rises by 45\% and \kam\ by 48\%.

Figure~\ref{fig:Ktopi} shows the K/$\pi$ ratio as a function of
\s\ both in pp (full symbols,
\cite{:2008ez,Alt:2005zq,Anticic:2010yg}) and in \pbar p (open
symbols,
\cite{Alexopoulos:1993wt,Alner:1985ra,Alexopoulos:1992ut})
collisions. For most energies, (\kap+\kam)/(\pip+\pim) is plotted,
but for some cases only neutral mesons were measured and
K$^{0}/\pi^0$ is used instead. The \pt-integrated
(\kap+\kam)/(\pip+\pim) ratio shows a slight increase from \s =
200 GeV (K/$\pi$ = $0.103\pm0.008$) to \s = 900 GeV
(K/$\pi$=$0.123\pm0.004\pm0.010$)~\cite{:2008ez}, yet consistent
within the error bars. The results at 7 TeV will show whether the
K/$\pi$ ratio keeps rising slowly as
 a function of \s{} or saturates.

\begin{figure}[th]
\resizebox{0.45\textwidth}{!}{%
\includegraphics{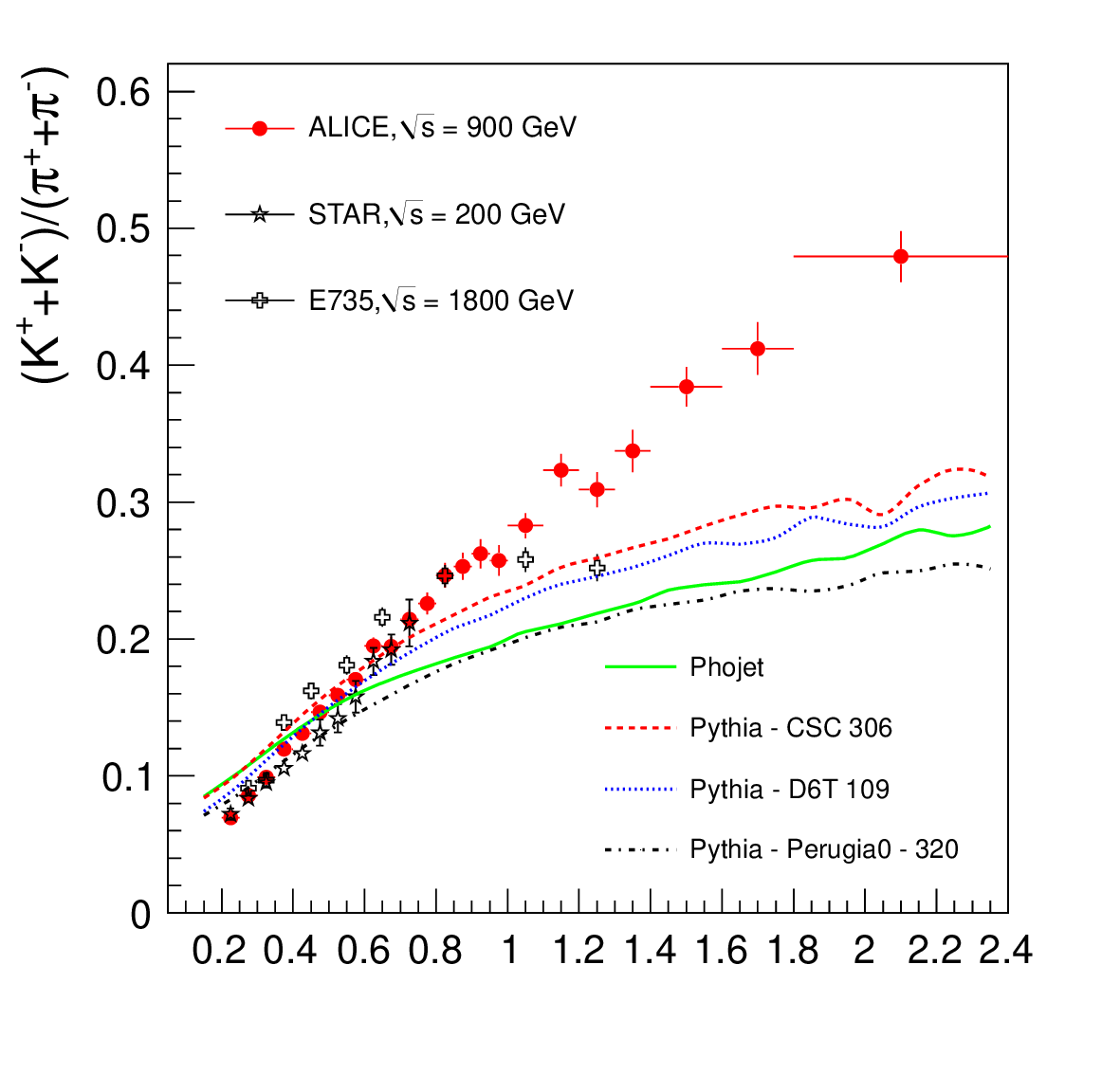}
}
\resizebox{0.45\textwidth}{!}{%
\includegraphics{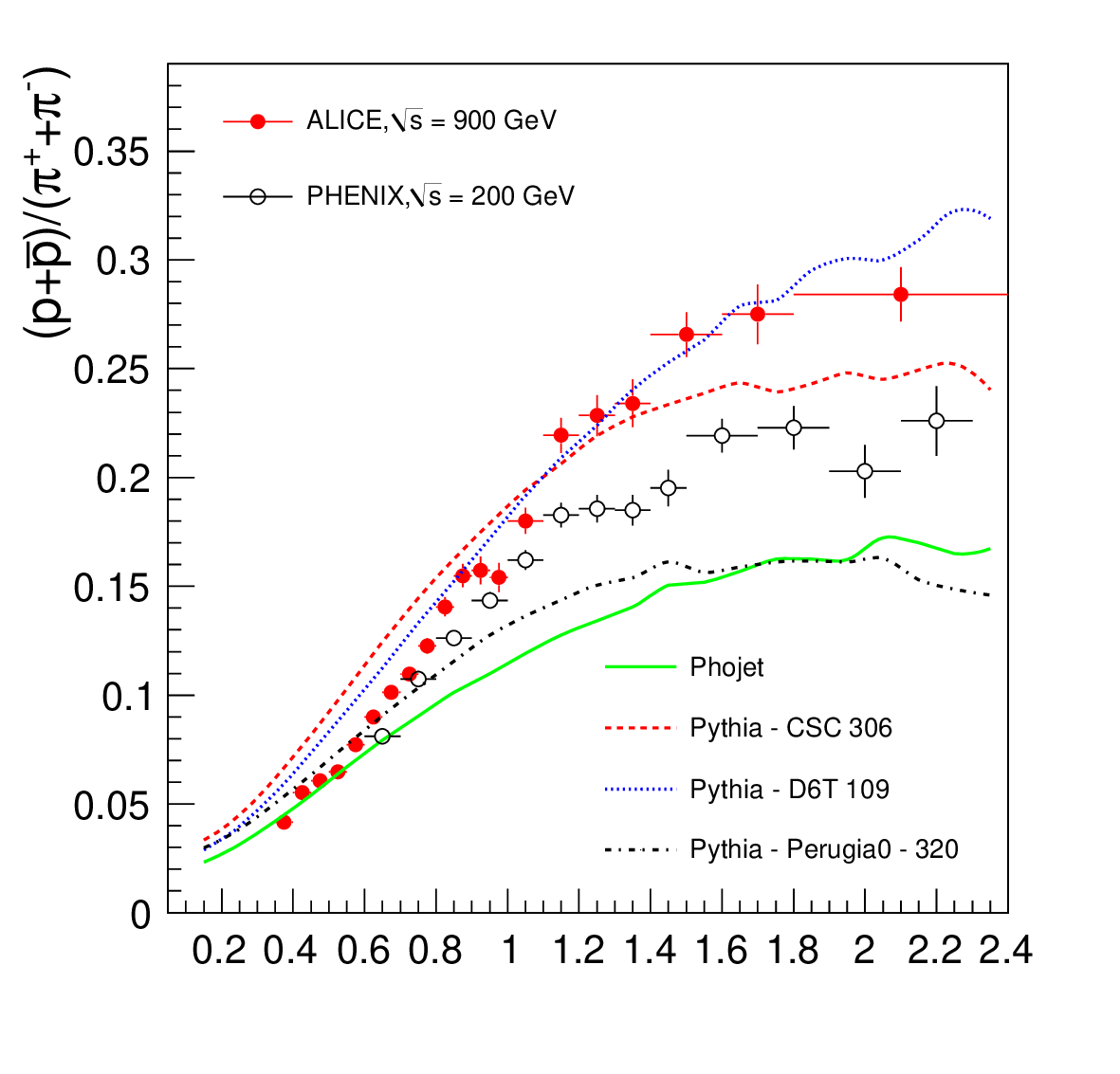}
} \caption{(Color online) Ratios of (\kap + \kam)/(\pip + \pim)
(upper panel) and (p + \pbar) / (\pip + \pim) (lower panel) as a
function of \pt\ from pp collisions at \s = 900 GeV (statistical
errors only). Values from the E735
Collaboration~\cite{Alexopoulos:1993wt} and the STAR
Collaboration~\cite{:2008ez}(upper part) and from the PHENIX
Collaboration~\cite{Adler:2006xd} (lower part) also are given. The
dashed and dotted curves refer to calculations using PYTHIA and
PHOJET at \s = 900 GeV.} \label{fig:ratios}
\end{figure}

Protons and antiprotons in Table~\ref{tab:levy} have been
corrected for feed down (mainly from $\Lambda$), while the results
from the STAR Collaboration are not. The proton spectra measured
by PHENIX, on the other hand, have a lower \pt-cut of 0.6~GeV/$c$.
This makes a direct comparison with RHIC data difficult.

Figure~\ref{fig:pythia-pos} shows a comparison of the measured
pion, kaon and proton spectra with several tunes of the PYTHIA
event generator~\cite{Sjostrand:2006za} and with
PHOJET~\cite{Engel:1995sb}. The PYTHIA CSC
306~\cite{Moraes:306AIN} tune provides a very poor description of
the particle spectra for all species. Similar deviations were
already seen for the unidentified charged hadron
spectra~\cite{Aamodt:2010my}. The other PYTHIA tunes,
Perugia0~\cite{perugia}  and D6T~\cite{D6T}, and PHOJET give a
reasonable description of the charged pion spectra, but show large
deviations in the kaon and proton spectra. The measured kaon
\pt-spectrum falls more slowly with increasing \pt\ than the event
generators predict. A similar trend is seen for the proton
spectra, except for PYTHIA tune D6T, which describes the proton
spectra reasonably well.

The upper panel of Figure~\ref{fig:ratios} shows the
\pt-depen\-dence of the K/$\pi$ and also the measurements by the
E735~\cite{Alexopoulos:1993wt} and STAR
Collaborations~\cite{:2008ez}. It can be seen that the observed
increase of K/$\pi$ with \pt\ does not depend strongly on
collision energy.

A comparison with event generators shows that at \pt $>$ 1.2
GeV/$c$, the measured K/$\pi$ ratio is larger than any of the
model predictions. It is interesting to note that while the
spectra in the CSC tune are much steeper than the other tunes, the
\pt-dependence of the K/$\pi$ ratio is very similar. In the
models, the amount of strangeness production depends on the
production ratios of gluons and the different quark flavours in
the hard scattering and on the strangeness suppression in the
string breaking. The latter could probably be tuned to better
describe the data. A similar disagreement between measured
strangeness production and PYTHIA predictions was found at RHIC
energies~\cite{Heinz:2007ci}.

In the bottom panel of Figure~\ref{fig:ratios}, the measured
p/$\pi$ ratio is compared to results at \s = 200 GeV from the
PHENIX Collaboration~\cite{Adler:2006xd}. Both measurements are
feed-down corrected. At low \pt, there is no energy-dependence of
the p$/\pi$ ratio visible, while at higher $\pt > 1$ GeV/$c$, the
p/$\pi$ ratio is larger at \s = 900 GeV than at \s = 200 GeV
energy.

Event generators seem to separate into two groups, one with high
p/$\pi$ ratio (PYTHIA CSC and D6T), which agree better with the
data and one group with a lower p/$\pi$ ratio (PHOJET and PYTHIA
Perugia0), which are clearly below the measured values. These
comparisons can be used for future tunes of baryon production in
the event generators.


\section{Summary}
\label{summary} We present the first analysis of transverse
momentum spectra of identified hadrons,  \pip, \pim, \kap, \kam,
p, and \pbar\ in pp collisions at \s\ = 900 GeV with the ALICE
detector. The identification has been performed using the \dedx\
of the inner silicon tracker, the \dedx\ in the gas of the TPC,
the kink
 topology of the decaying kaons inside the TPC and
the time-of-flight information from TOF. The combination of these
techniques allows us to cover a broad range of momentum.

Agreement in the K/$\pi$ ratio is seen when comparing to \pbar p
collisions at the Tevatron and Sp\pbar S. Comparing our results
with a similar measurement from the STAR Collaboration using pp
collisions at \s = 200 GeV the shape of the spectra shows an
increase of the hard component, but we observe only a slight
increase of the mean \pt-values. Whether the fraction of strange
to non-strange particles rises
with increasing \s\ remains open until data at 7 TeV become available.\\

\clearpage

\noindent{\bf Acknowledgements}\\

The ALICE collaboration would like to thank all its engineers and technicians for their invaluable contributions to the construction of the experiment and the CERN accelerator teams for the outstanding performance of the LHC complex.
\\
The ALICE collaboration acknowledges the following funding agencies for their support in building and
running the ALICE detector:
 \\
Department of Science and Technology, South Africa;
 \\
Calouste Gulbenkian Foundation from Lisbon and Swiss Fonds Kidagan, Armenia;
 \\
Conselho Nacional de Desenvolvimento Cient\'{\i}fico e Tecnol\'{o}gico (CNPq), Financiadora de Estudos e Projetos (FINEP),
Funda\c{c}\~{a}o de Amparo \`{a} Pesquisa do Estado de S\~{a}o Paulo (FAPESP);
 \\
National Natural Science Foundation of China (NSFC), the Chinese Ministry of Education (CMOE)
and the Ministry of Science and Technology of China (MSTC);
 \\
Ministry of Education and Youth of the Czech Republic;
 \\
Danish Natural Science Research Council, the Carlsberg Foundation and the Danish National Research Foundation;
 \\
The European Research Council under the European Community's Seventh Framework Programme;
 \\
Helsinki Institute of Physics and the Academy of Finland;
 \\
French CNRS-IN2P3, the `Region Pays de Loire', `Region Alsace', `Region Auvergne' and CEA, France;
 \\
German BMBF and the Helmholtz Association;
 \\
Hungarian OTKA and National Office for Research and Technology (NKTH);
 \\
Department of Atomic Energy and Department of Science and Technology of the Government of India;
 \\
Istituto Nazionale di Fisica Nucleare (INFN) of Italy;
 \\
MEXT Grant-in-Aid for Specially Promoted Research, Ja\-pan;
 \\
Joint Institute for Nuclear Research, Dubna;
 \\
National Research Foundation of Korea (NRF);
 \\
CONACYT, DGAPA, M\'{e}xico, ALFA-EC and the HELEN Program (High-Energy physics Latin-American--European Network);
 \\
Stichting voor Fundamenteel Onderzoek der Materie (FOM) and the Nederlandse Organisatie voor Wetenschappelijk Onderzoek (NWO), Netherlands;
 \\
Research Council of Norway (NFR);
 \\
Polish Ministry of Science and Higher Education;
 \\
National Authority for Scientific Research - NASR (Autoritatea Na\c{t}ional\u{a} pentru Cercetare \c{S}tiin\c{t}ific\u{a} - ANCS);
 \\
Federal Agency of Science of the Ministry of Education and Science of Russian Federation, International Science and
Technology Center, Russian Academy of Sciences, Russian Federal Agency of Atomic Energy, Russian Federal Agency for Science and Innovations and CERN-INTAS;
 \\
Ministry of Education of Slovakia;
 \\
CIEMAT, EELA, Ministerio de Educaci\'{o}n y Ciencia of Spain, Xunta de Galicia (Conseller\'{\i}a de Educaci\'{o}n),
CEA\-DEN, Cubaenerg\'{\i}a, Cuba, and IAEA (International Atomic Energy Agency);
 \\
Swedish Reseach Council (VR) and Knut $\&$ Alice Wallenberg Foundation (KAW);
 \\
Ukraine Ministry of Education and Science;
 \\
United Kingdom Science and Technology Facilities Council (STFC);
 \\
The United States Department of Energy, the United States National
Science Foundation, the State of Texas, and the State of Ohio.

\bibliographystyle{epjc}
\bibliography{id}

\end{document}